\documentclass{aa}
\usepackage{multirow}
\usepackage{graphicx}
\usepackage{txfonts}
\usepackage{natbib}
\usepackage{natbib}

\begin{document}

\title{Stellar longitudinal magnetic field determination through multi-Zeeman signatures}

\author{ 
     {J.C. Ram\'{\i}rez V\'elez}\inst{1}
\and {M.J. Stift}\inst{2}
\and {S. G. Navarro}\inst{3} 
\and {J.P. C\'ordova}\inst{3}
\and {L. Sabin}\inst{4}
\and { A. Ruelas-Mayorga}\inst{1}
}
\institute{Instituto de Astronomia - Universidad Nacional Autonoma de Mexico, 
           Apdo. Postal 70-264, 04510, D. F., Mexico,  \email{jramirez@astro.unam.mx}
\and Armagh Observatory, College Hill, Armagh BT61 9DG. Northern Ireland, \email{stiftm7@gmail.com}
\and Universidad de Guadalajara, Av. Vallarta 2602, Arcos Vallarta, Guadalajara, Jalisco, Mexico, \email{silvana@astro.iam.udg.mx, jcordova@astro.unam.mx}
\and Instituto de Astronomia - Universidad Nacional Autonoma de Mexico, Apdo. Postal 877,
     22860, Ensenada, B. C., Mexico. 
     \email{lsabin@astrosen.unam.mx}
}

\offprints{jramirez@astro.unam.mx}
\date{}

\abstract
{A lot of effort has been put into the detection and determination of stellar magnetic 
fields using the spectral signal obtained from the combination of hundreds or thousands 
of individual lines, an approach known as ``multi-line techniques''.  However, so 
far  most of multi-line techniques developed that retrieve stellar mean longitudinal 
magnetic fields  recourse to sometimes heavy simplifications
concerning line shapes and Zeeman splittings.}
{  To determine stellar longitudinal magnetic fields by means of the Principal
 Components Analysis and Zeeman Doppler Imaging (PCA-ZDI) multi-line technique,
 based on accurate polarised spectral line synthesis.} 
{In this paper we present the methodology to perform inversions of profiles obtained with 
 PCA-ZDI.}
{  Inversions with various magnetic geometries, field strengths and rotational
 velocities show that we can correctly determine the  effective longitudinal magnetic field in
 stars using the PCA-ZDI method.} 
{}


\titlerunning{Inversions of synthetic PCA-ZDI profiles}
\authorrunning{Ram\' irez V\'elez  et al.\ \ }
\maketitle

\section{Introduction}

It is well known that the magnetic solar type stars host weak longitudinal fields, 
typically of the order of a few tens of Gauss (e.g. \citealt{Marsdenetal2014}). For 
such field strengths, Stokes $V$ circular polarisation signatures of individual 
spectral lines are generally well below the noise level. The use of so-called 
multi-line techniques makes it possible to overcome this problem through the 
``addition'' of multiple individual lines in Doppler space, resulting in a ``mean'' 
circular polarisation profile, the so-called Multi-Zeeman-Signature (MZS) which 
greatly increases the signal to noise ratio. Starting with the pioneering work of
\citet{SemelLi1996}, different techniques have been developed over the following 
two decades for the establishment of Multi-Zeeman-Signatures. The most popular 
appears to be the Least Squares Deconvolution (LSD) technique first described in 
\citet{Donatietal1997}. Two assumptions underlie the LSD technique. The first one 
states that the local circular polarisation profiles to be added are all of 
similar shape; the second postulates that the Zeeman broadening is very small 
compared to the thermal Doppler broadening, such that the weak field approximation 
can safely be applied to the Stokes profiles. As a consequence, the  coupled 
system of  equations of polarised radiative transfer do not have to be formally 
solved.  Instead, under a perturbative scheme the system of equations can 
be partially decoupled, permitting to find an analytical solution: 
to first order the circular polarisation profile is proportional to the 
first derivative of the intensity, and to second order, the linear 
polarised profiles are proportional to the second derivative of the intensity 
(e.g. \citealt{LandiLandolfi2004}). In recent times, LSD and the weak field 
approximation (WFA) have been widely employed at different levels of 
sophistication with the ultimate goal of precisely measuring weak stellar 
magnetic fields on the basis of MZSs (\citealt{KochukhovMaPi2010}, 
\citealt{Kochukhov2015}, \citealt{MartinezAsensio2012}, 
\citealt{Martinezetal2012}, \citealt{CarrollStr2014}, \citealt{AsensioPet2015}). 

An alternative  technique based on the Principal Components Analysis (PCA) was 
proposed by \citet{Martinezetal2008}. In this work, the MZS is derived by means 
of the addition of many lines in Doppler space (as in LSD) but a denoising 
procedure, the filtering of uncorrelated noise, is applied to individual lines to 
increase the signal to noise ratio in the final MZS. It is an important feature of 
this technique that similarity between the individual polarised circular line 
profiles is not required; neither is it necessary to invoke the WFA. The validity 
of this robust technique for the analysis of magnetic fields has been proved 
with numerical simulations but the method has not yet been applied to the 
quantitative {\em measurement} of field strengths.

In fact, most current techniques devoted to the analysis of stellar magnetic fields 
by means of MZS are avoiding spectral line synthesis on account of the computing
resources required in view of the wide spectral ranges involved (thousands of 
Angstroms). To contribute to remedy this situation, we will present the 
basis of a novel technique for the determination of stellar longitudinal magnetic 
fields, a method that is based on spectral line synthesis incorporating detailed 
polarised radiative transfer.

\section{``Solar'' and stellar Stokes profiles}

In this section we shall use the Stokes profiles obtained from a given single 
point on the surface of the star -- we call them ``solar'' profiles -- to 
derive the Stokes profiles integrated over the entire visible hemisphere of 
the star; we shall in what follows denote the latter as ``stellar'' profiles. 
To start with, we first consider a star with zero rotational velocity, but 
later on, this constraint will be removed.

A number of parameters have to be specified for the calculation of Stokes 
profiles of magnetic stars. In the present case, that we employ the eccentric 
dipole oblique rotator model \citep{StiftSt1975}, these are the effective temperature 
$T_{\rm eff}$, the surface gravity $\log g$, the metallicity [$M/H$], the 
projected rotational velocity of the star $v\,\sin i$, the macro-turbulent 
($v_{\rm turb}$) and the micro-turbulent ($\xi$) velocities, the position of the 
magnetic dipole inside the star -- given by the two coordinates $x_2$ and $x_3$ 
-- and the dipole moment $\mathbf{m}$. In addition, it is necessary to specify 
the orientation of the magnetic axis with respect to the rotation axis with the 
help of the 3 Eulerian angles $\alpha, \beta, \gamma$, and the orientation of 
the rotational axis with respect to the line of sight (LOS), i.e. the inclination 
angle $i$. See \citet{StiftSt1975} for details.  For the synthesis of 
the Stokes spectra we have employed {\sc Cossam} (\citealt{Stift2000},
\citealt{StiftStLeCo2012}), an LTE code that calculates the four Stokes 
parameters by detailed opacity sampling and by accurately solving the coupled 
equations of polarised radiative transfer.

\subsection{Spatial grids}

The so-called effective magnetic field $H_{\rm eff}$ -- which represents a weighted 
average of the longitudinal component of the magnetic field vector over the visible 
hemisphere of the star -- depends on the wavelength range covered by the spectral 
lines, on their atomic properties, on their strength, and on the limb darkening 
\citep{Stift1986b}. The spatial distribution over the stellar surface of the quadrature 
points may also play a role if the spatial resolution is too low in the presence of 
rotation, pulsation and/or strong magnetic fields. Therefore one has to ensure the 
establishment of a spatial grid figuring a decent number of well distributed points. 
We may distinguish two main types of grids, viz. fixed grids and adaptive grids. The 
{\sc Cossam} code provides both grid types. Fixed grids can be co-rotating as found in
Doppler mapping (see e.g. \citealt{VogtVoPeHa1987}) or observer-centred. For the former 
one tries to create surface elements of almost identical size in bands equidistant
in latitude, whereas for the latter the points are arranged in concentric rings
about the centre of the visible stellar disk. The concentric rings are very 
popular because even a quite small number of such rings usually proves sufficient 
for the spatial integration required to obtain quantities like the mean 
longitudinal magnetic field $H_{\rm eff}$ or the mean field modulus $H_{\rm s}$.

Polarised line synthesis in fast rotating stars can constitute a challenge for 
these simple fixed grids which may lead to a uncomfortably large number of
points. Adaptive grids are better in general, representing non-uniform 
distributions with no pre-determined distances between the individual points. 
The central idea behind adaptive grids in line synthesis is to ensure that the 
difference in monochromatic line opacity between two adjacent points should not 
exceed a certain percentage -- refer to \citet{Stift1985} for details. Let us for 
example consider the sampling of the central opacity of a line at a given point. 
If the grid resolution is not high enough, it may happen that at a neighbouring 
point the line opacity is already negligible as a result from differences in the 
magnetic field strength and angle, or from the Doppler effect due to rotation 
or pulsation. The integrated line intensity would thus reach its minimum at 
one point, whereas at the neighbouring point there would be nothing but the 
continuum. However sophisticated the quadrature scheme, the integrated line 
profile would always be in error for such a sparse grid.

In {\sc Cossam} the density of points can be adjusted through five parameters.
As a proxy for the limit in the relative percentage change in opacity between 
neighbouring quadrature points one takes a fraction of the Doppler width of
the opacity profile of a typical metal line ($R_{\rm dop}$). In the 
polarised radiative radiation equations (see \citealt{AlecianAlSt2004})
two angles enter: the angle between magnetic field vector and the LOS on the
one hand, and the azimuth angle defining Stokes $Q$ and $U$ on the other 
hand; they are taken care of by $R_{\rm cos}$ and $R_{\rm azi}$ respectively. 
There remains the continuum intensity $R_{\rm flux}$ which may not be
neglected in the establishment of the spatial grid. Finally one has to
ensure the monotonicity of the resolution function in those cases where
the Doppler shift between adjacent points is zero and where the field angles 
are identical. For this purpose a small empirical constant $\Delta$ is added 
to the resolution function.
Adaptive grids are very useful thanks to the optimal placement of the
quadrature points and their reduced number compared to fixed grids. This 
can significantly reduce the time needed to calculate Stokes profiles 
in strongly magnetic and/or non-radially pulsating stars \citep{Fensl1995}.

\begin{figure*}
\hspace{0.5cm}\includegraphics[width=7.5cm]{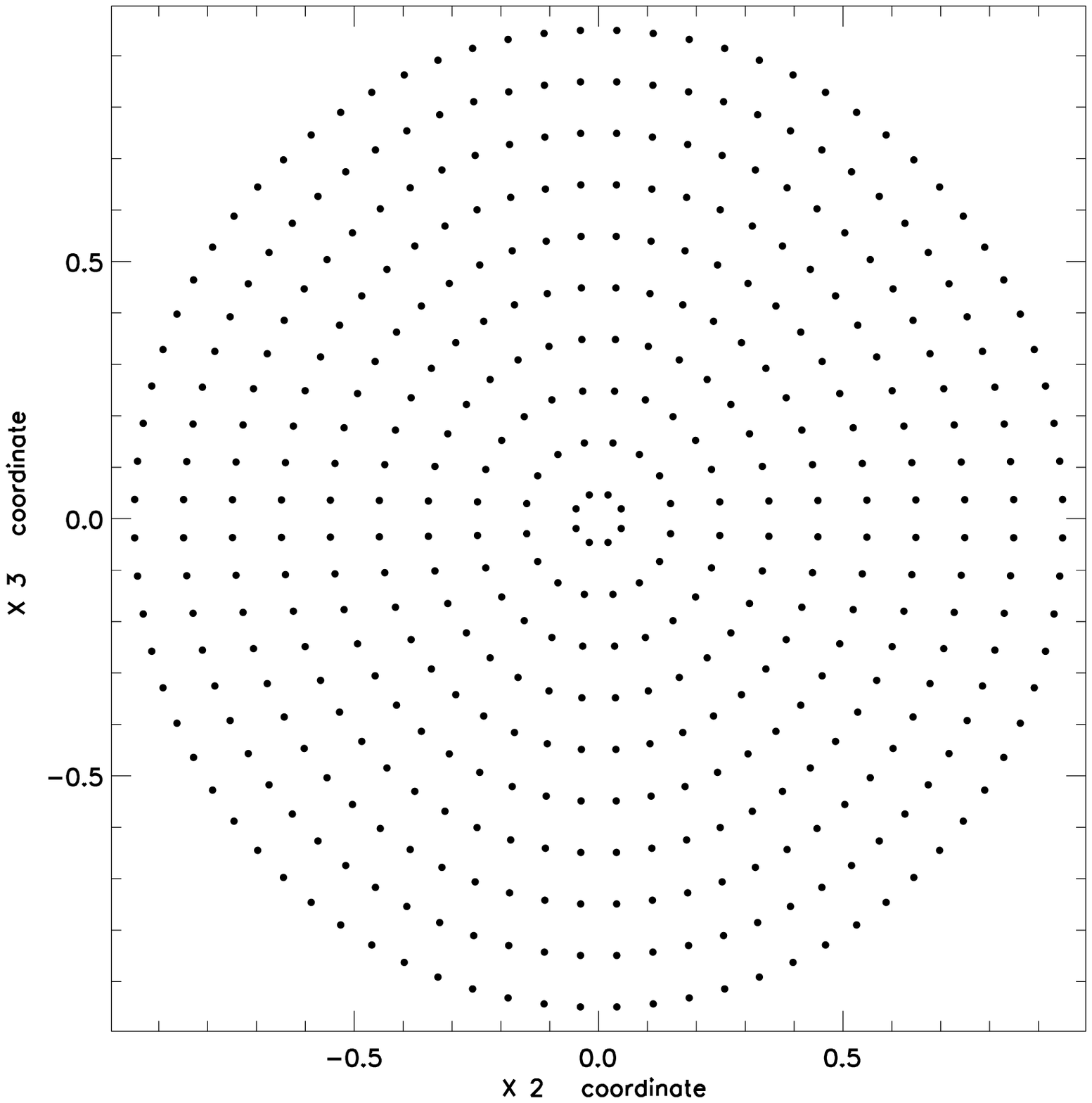}
\hspace{2.0cm}\includegraphics[width=7.5cm]{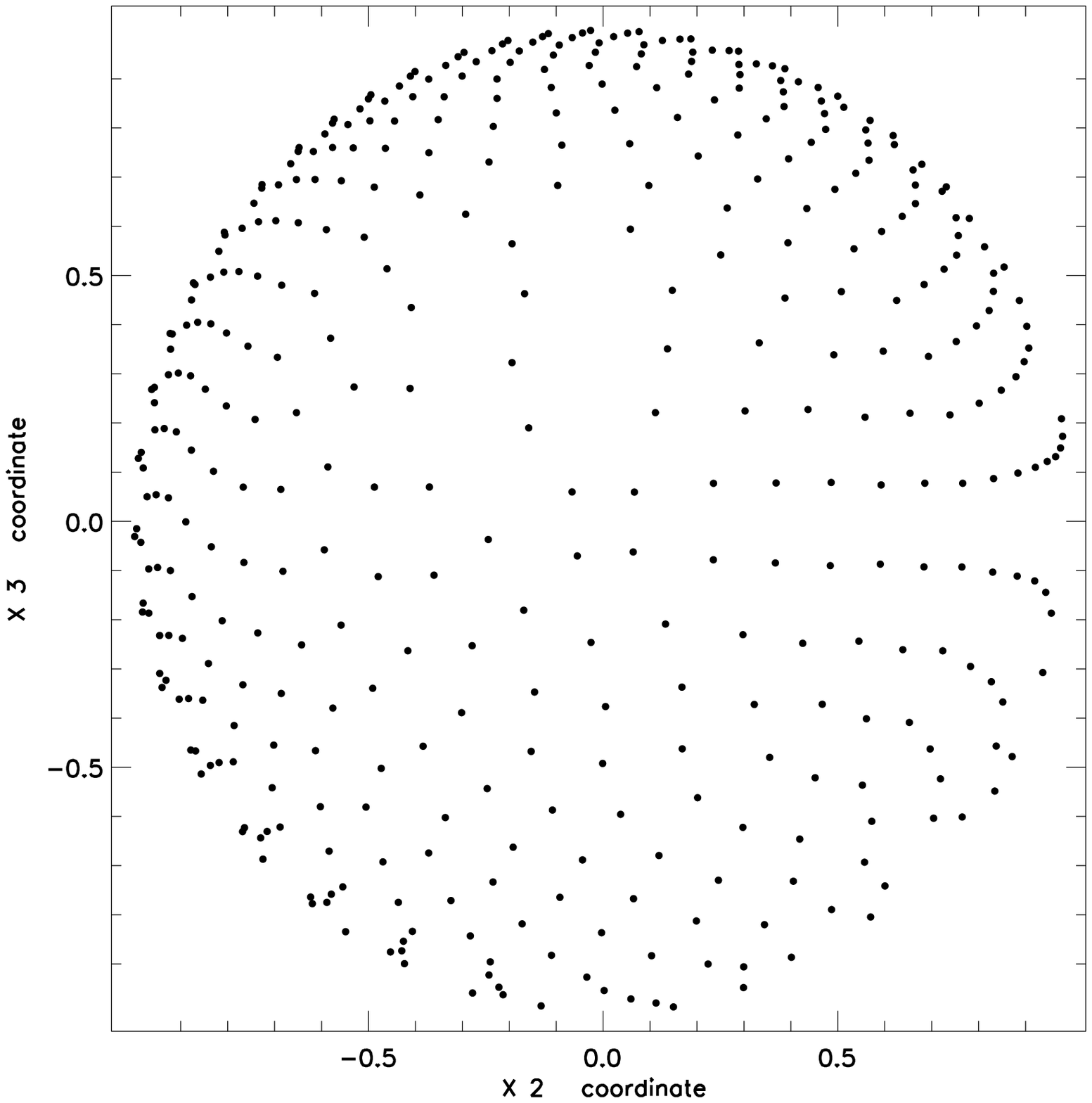}
\caption{Examples of quadrature point distributions for a rigid (left) and an 
   adaptive grid (right). For illustrative purposes, the number of quadrature points
   is nearly the same in both cases.}
\label{fig:grids}
\end{figure*}

\begin{figure*}
\hspace{-0.6cm}\includegraphics[width=10cm]{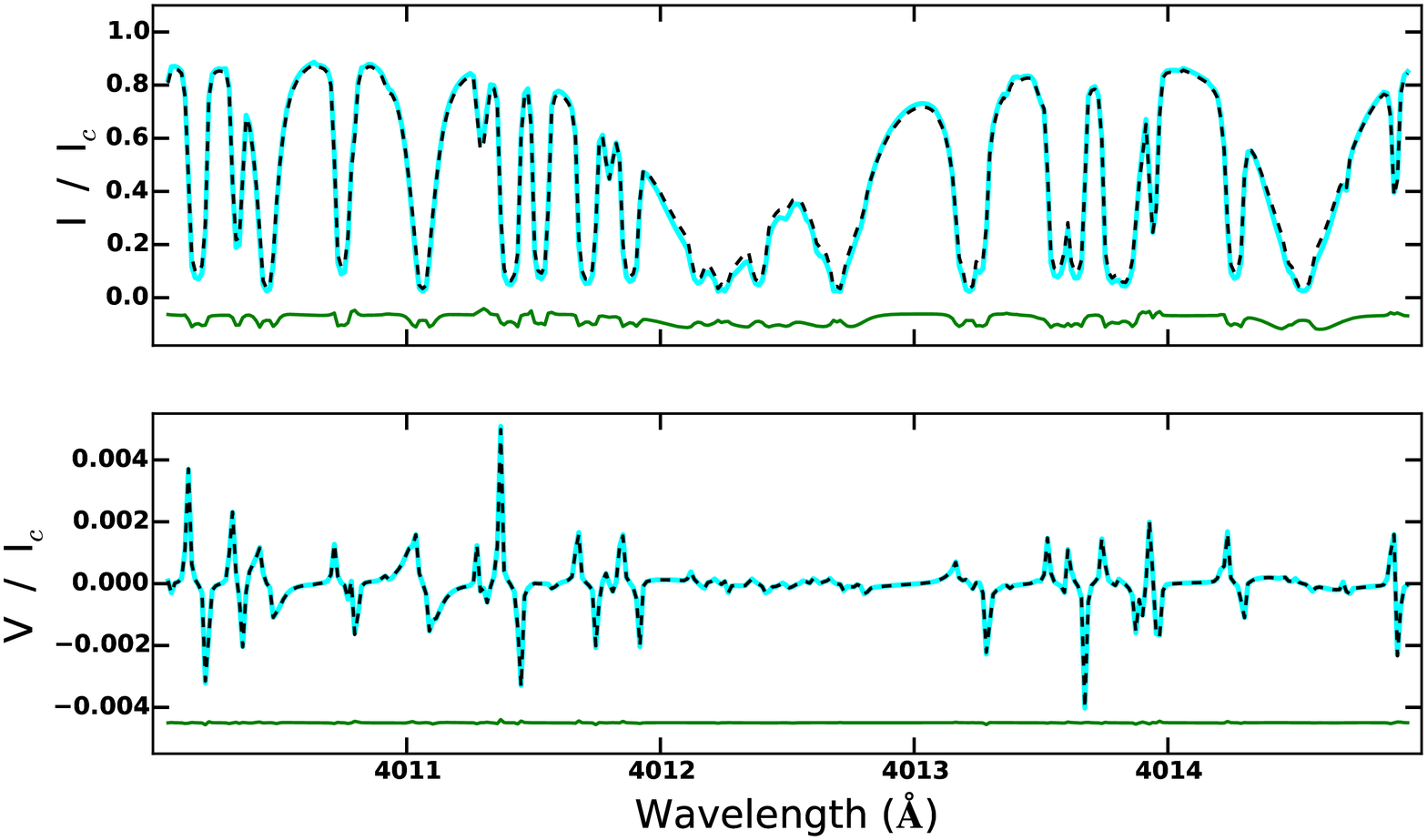}
\hspace{-0.3cm}\includegraphics[width=10cm]{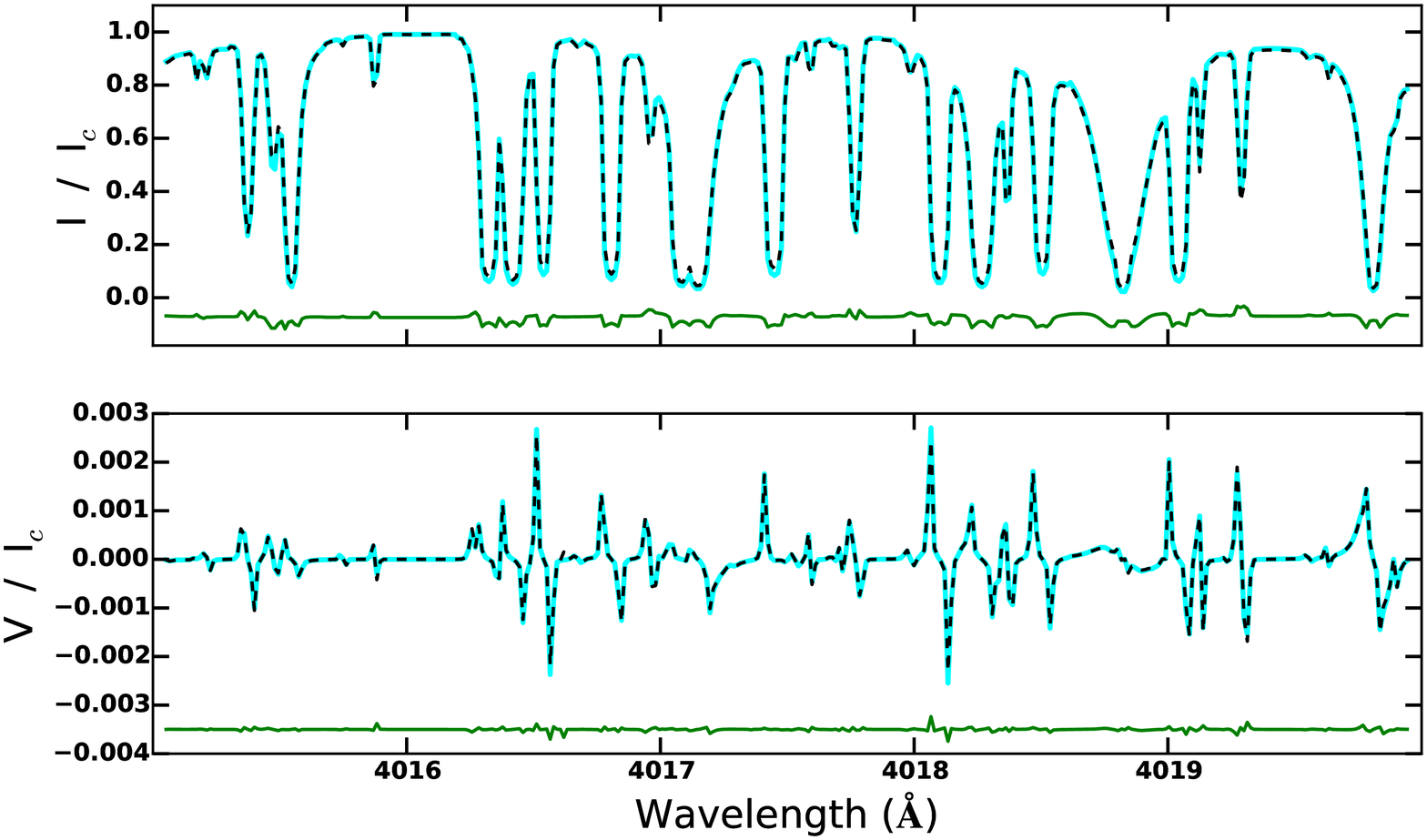}
\caption{Comparison between ``solar'' and ``stellar'' (integrated) Stokes
   profiles. The dashed black line pertains to the integrated Stokes profiles, 
   the solid line in light colour to the ``solar'' case. The difference in the
   profiles, shown at the bottom, has been shifted for clarity. The effective
   longitudinal field strengths are 7.65\,G (left panel) and 5.78\,G (right panel).}
\label{fig:comp}
\end{figure*}

\subsection{``Stellar'' (integrated) versus ``solar'' (point source) profiles}

We adopted two particular magnetic field configurations and spatial grids for the
demonstration of how well stellar Stokes profiles are approximated by ``solar''
profiles. In the first case, we assumed an oblique rotator seen equator-on, with
the magnetic dipole located at the centre of the star and its axis perpendicular 
to the rotational axis (Model\,1 in Table \ref{tab:magmod}). At phase zero, the 
magnetic pole is at the centre of the visible disk, the field is thus symmetric 
around the line of sight. In this case, we employed a fixed grid with 10 rings 
where the innermost one features 8 surface elements, for a total of 440 points, 
shown in the left panel of Fig.\,\ref{fig:grids}. For the second magnetic 
configuration, labelled Model\,2 in Table\,\ref{tab:magmod}, we chose a
decentred tilted dipole, resulting in a highly
inhomogeneous distribution of field strengths and angles over the visible
hemisphere. With the adaptive grid parameters $\Delta = 0.5$, $R_{\rm dop} = 0.5$, 
$R_{\rm cos} = 0.3$, $R_{\rm azi} = 0.3$, and $R_{\rm flux} = 0.3$ we arrived at a total 
of 406 points distributed as shown in the right panel of Fig.\,\ref{fig:grids}. 

\begin{table}
\caption{Magnetic models used for the calculations of the Stokes profiles. 
For the definition of the model parameters see the text.}
\label{tab:magmod}
 \begin{tabular}{|c|cccccc|}
  \hline
     &  i ($^{\circ}$) & $\alpha$ ($^{\circ}$) & $\beta$ ($^{\circ}$) 
     & $\gamma$ ($^{\circ}$) & $x_2$ ($R_{\star}$) & $x_3$ ($R_{\star}$) \\
  \hline
 Model 1 & 90 & -90  &   0 &   0  & 0.0 &   0.0 \\  
 Model 2 & 15 &  10  &  23 & -34  & 0.2 &  -0.3 \\  
 \hline
 \end{tabular}
 \end{table}

For the synthesis of the Stokes profiles with {\sc Cossam} we selected from the Atlas9 
grid \citep{CastelliKurucz2004} an atmospheric model with $T_{\rm eff} = 5\,500\,K$, 
$\log g = 4.0$, solar metallicity, zero macro-turbulence ($v_{\rm turb} = 0.0$) and a 
micro-turbulent velocity of $\xi = 2.0$\,km\,s$^{-1}$. We then adopted the same dipole 
moment ($\boldmath{m} = 10$\,G) for both models, giving an effective longitudinal 
field $H_{\rm eff}$ of 7.65 and 5.78\,G respectively (for the wavelength range from 4000 
to 4050\,{\AA};  please be aware that $H_{\rm eff}$ depends on the wavelength range 
considered). We also calculated Stokes profiles for the ``solar'' case, having the 
magnetic field vector at the centre of the visible disk point towards the observer and 
setting the value of the field strength  equal to the value of  $H_{\rm eff}$. 
In order to compare the ``solar'' to the ``stellar'' (integrated) Stokes profiles, we 
finally normalised them to the continuum.

In what follows we shall discuss only the Stokes profiles in intensity $I$ and in
circular polarisation $V$ given that in this study we are solely interested in the 
determination of the longitudinal magnetic field. In Fig.\,\ref{fig:comp} we show 
the stellar (integrated) Stokes profiles (dashed black line) over a narrow interval 
of 5\,{\AA} and the respective ``solar'' Stokes profiles (solid line in light 
colour). The difference between the stellar and the ``solar'' profiles is shown 
at the bottom of each panel.  The Stokes profiles in the left (right) side 
panels were calculated with the grid shown in the left (right) side panel of 
Fig.\,\ref{fig:grids}. 

From straightforward visual inspection it is clear that the ``solar'' profiles
very closely reproduce the Stokes profiles obtained by integration of the local
profiles over the visible hemisphere (there are more than 400 points for the
two spatial grids). We quantify this result by evaluating the mean absolute
error (MAE) of the difference between the ``solar'' and the ``stellar'' $V$
profiles (lower panels of Fig.\,\ref{fig:comp}). Additionally, in order to
test how well the effective longitudinal magnetic field can be approximated,
we repeat the same procedure for the ``solar'' profiles but with increments
of $\pm$ 0.5\,G in field strength. The results are listed in Table \ref{grids_0km}.

\begin{table}
 \centering
\caption{Comparison of $V$ profiles for the two grids in Fig. \ref{fig:grids}.}
\label{grids_0km}
\begin{tabular}{|c|cll|}
\hline
Fixed      & "Stellar" field    & "Solar"            & MAE for \\
grid       & $H_{\rm eff}$ (G)  & field (G)          & $V$ profiles\\
 \hline
          & 7.65        &    8.65             & $4.0 \times 10^{-5}$ \\
Magnetic  & 7.65        &    8.15             & $2.0 \times 10^{-5}$ \\
          & 7.65        &    7.65             & $1.0 \times 10^{-5}$ \\
Model 1   & 7.65        &    7.15             & $2.1 \times 10^{-5}$ \\
          & 7.65        &    6.65             & $4.0 \times 10^{-5}$ \\
 \hline  
 \end{tabular}
 
 \vspace{1cm}\begin{tabular}{|c|cll|}

\hline
Adaptive    & "Stellar" field     & "Solar"            & MAE for  \\
grid        & $H_{\rm eff}$ (G)   & field  (G)         & $V$ profiles\\
 \hline
          & 5.78        &    6.78             & $4.1 \times 10^{-5}$ \\
Magnetic  & 5.78        &    6.28             & $2.1 \times 10^{-5}$ \\
          & 5.78        &    5.78             & $1.6 \times 10^{-5}$ \\
Model 2   & 5.78        &    5.28             & $2.1 \times 10^{-5}$ \\
          & 5.78        &    4.78             & $4.1 \times 10^{-5}$ \\
 \hline 
  \end{tabular}
\end{table}

We note that for both grids the ``solar'' fit to the stellar $V$ Stokes
profiles is highly accurate. Using the value of $H_{\rm eff}$ for input in
a ``solar'' type polarised line synthesis, one obtains a near perfect fit;
even a difference of a mere $\pm 0.5$\,G already leads to a less satisfactory
result.

\subsection{Including stellar rotation}
\label{sec2.3}

In the preceding section we showed that ``solar''  profiles can be used instead of
stellar (integrated) profiles for the analysis of stellar magnetic fields in
the case of negligible rotational broadening. In this section we shall discuss what
happens if we include the effects of rotational broadening.

In order to establish a proper broadening function (BF) that can be applied to the
``solar'' profiles $S_{\rm p}$, producing a correct fit to the stellar rotationally
broadened profiles $S_{\rm rot}$, one has to solve the equation
\begin{equation}
 S_{\rm rot}(\lambda) = BF(\lambda) * S_{\rm p}(\lambda),
 \label{ecBF}
\end{equation}
where the symbol ``$*$'' denotes the convolution of the two functions and $S$ represents
any of the Stokes $IQUV$ parameters. Two approaches are possible to obtain the BF,
viz. the cross-correlation (CC) or the Fourier deconvolution methods respectively.
The former produces a BF which acts more like a proxy, while the latter provides a
more accurate broadening function. \citet{Rucinski1999} showed the advantages of using
the Fourier deconvolution rather than CC function, but, more importantly, also
showed how the deconvolution approach can be transformed into a linear system of
equations which in turn can be solved through Singular Value Decomposition (SVD).

Let $m$ be the number of wavelength points of both ``solar'' and stellar 
profiles, and let $n$ be the number of points (window length) of the broadening 
function. The central idea is to use $S_{\rm p}$ to construct a matrix $\hat{M}$ of 
dimensions $(m-n) \times n$. Each column in matrix $\hat{M}$   is composed of a 
segment (of length $m-n$) of the profiles $S_{\rm p}$ but shifted vertically by one 
wavelength element. Let $S_{\rm p}^*$ denote the segment of interest containing the 
first $m-n$ elements of $S_{\rm p}$\,. The column arrangement is as follows: the 
last column of $\hat{M}$ is represented by $S_{\rm p}^*$\,, the penultimate by 
$S_{\rm p}^*$ shifted by one wavelength element, and so on. Finally, for the proper 
handling of the edges of the profiles $S_{\rm rot}$\,, a reduction from $m$  to 
$ m-n $ elements is applied. This allows to remove $n/2$ points from each side 
of $S_{\rm rot}$ -- more details can be found in \citet{Rucinski1992} and in
\citet{Rucinski1999}. By applying this rearrangement, Eq.\,(\ref{ecBF}) is 
transformed to a linear system of equations:
 
 \begin{equation}
\hat{M}(\lambda) \ BF(\lambda) = S_{\rm rot}(\lambda).
 \label{ecM}
\end{equation}

This system of equations is solved applying SVD to $\hat{M}$, from which we 
obtain the orthonormal matrices $\hat{U}$ and $\hat{V}$ and the vector $w$, 
with respective dimensions of $(m-n) \times n$, $n \times n$, and $n$ (see 
e.g. \citealt{GolubvanLoan1996}). With the elements of the vector $w$, it is 
possible to define the square matrix $\hat{W_1}$ (with dimensions $n \times n$) 
which is different from zero only in the diagonal elements as $W_1(i,i)=w(i)$. It 
is then possible to write the matrix $\hat{M}$ as the product of these 3 matrices
\begin{equation}
\hat{M}(\lambda) = \hat{U}(\lambda)\hat{W_1}(\lambda)\hat{V^{\rm T}}(\lambda).
\end{equation}

\begin{figure}
\hspace{-0.5cm}\includegraphics[width=10cm]{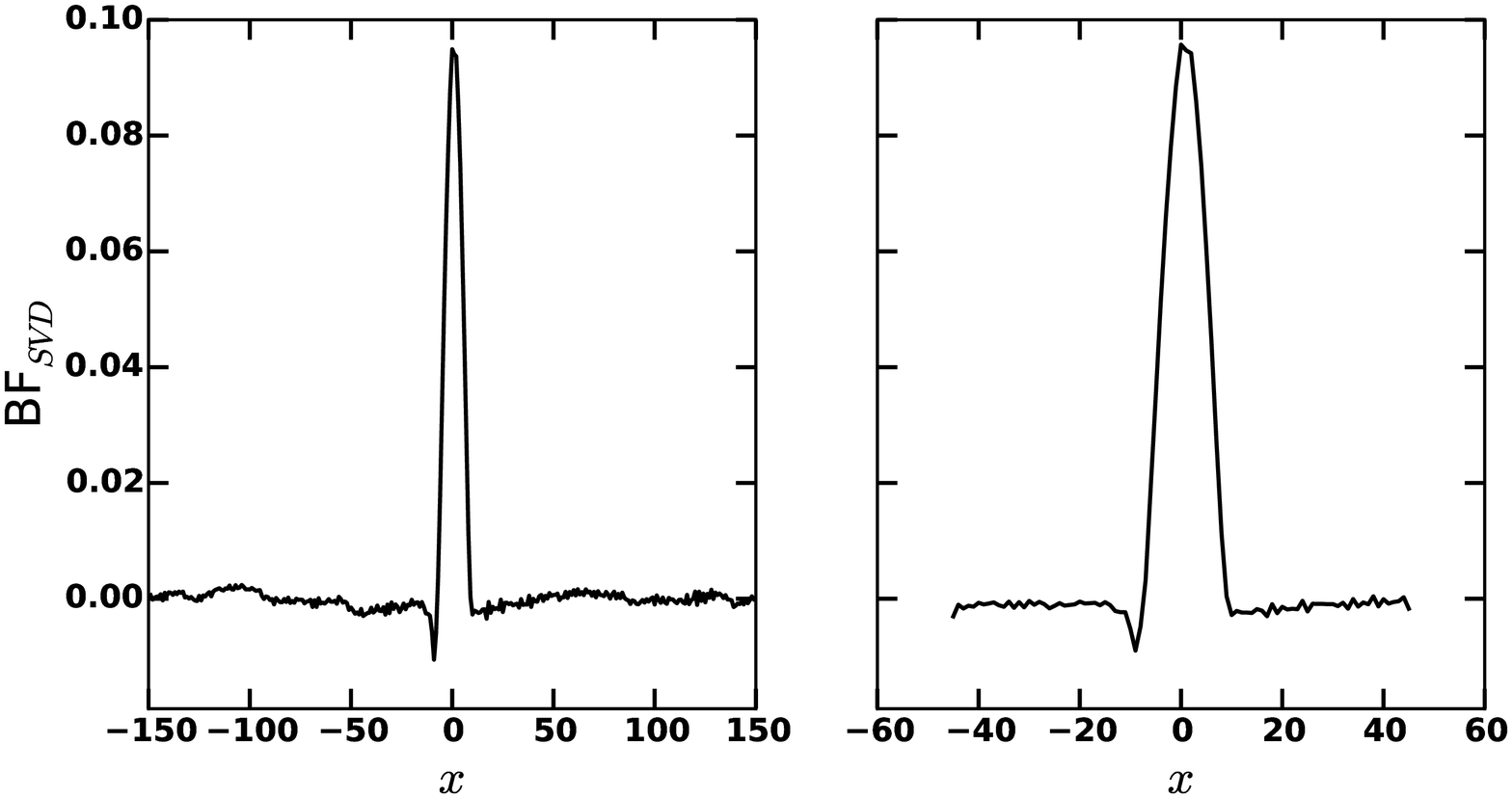}
\includegraphics[width=9cm]{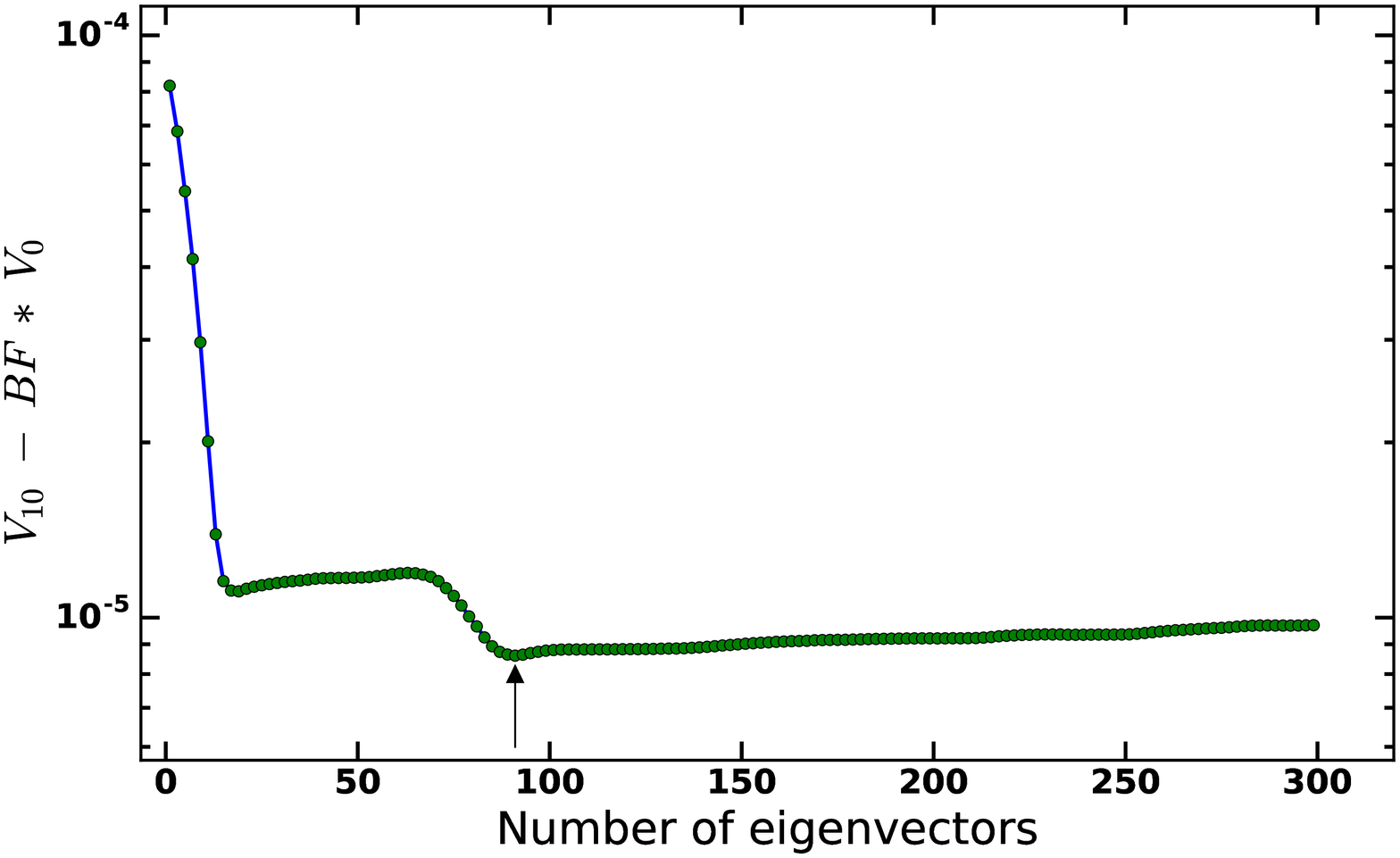}
\caption{Upper left panel: a wide broadening function using
a spectral window of 301 points. Upper right panel: optimal BF$(x)$ with
91 spectral points. The independent variable $x$ has been substituted for
$\lambda$ in order to emphasise that the sampling rate is at constant velocity
steps (1.0\,km\,s$^{-1}$ in our case). Lower panel: comparison of $V$ Stokes
parameters as a function of the number of eigenvectors included in the BF.
The optimum fit achieved with 91 eigenvectors is indicated by an arrow.}
\label{fig:BF}
\end{figure}

Let us now define $\hat{W}$ as the diagonal matrix whose elements are the inverse 
of the diagonal elements of $\hat{W_1}$, viz. $W(i,i)=1/w(i)$. Finally, taking
the inverse of the orthonormal matrices $U$ and $V$ (their transposes), it becomes
possible to recover the broadening function as (omitting the dependence on $\lambda$):
\begin{equation}
 BF = \hat{V} \hat{W} \hat{U^T}S_{\rm rot}.
 \label{ec_solBF}
\end{equation}

SVD has many interesting properties, one of them being that the eigenvectors in
$\hat{U}$ and $\hat{V}$ are arranged in order of importance. This fact allows a
reduction of dimensionality without loss of useful information. In practice, we
implement the dimensionality reduction by storing in the diagonal of the matrix
$\hat{W}$ a number $l < n$ of elements. To illustrate how to find the optimal
number $l$, we use the circular Stokes profiles shown in the right panels of
Fig.\,\ref{fig:comp}. Let $S_{0}$ denote the ``solar'' Stokes profiles and
$S_{10}$ the stellar (integrated) Stokes profiles calculated with the adaptive
grid and $v\,\sin i = 10\,{\rm km}\,{\rm s}^{-1}$. We first calculate the Stokes 
profiles with only one eigenvector ($l = 1$) using Eq.\,(\ref{ecBF}),
i.e. $S_{10} = {\rm BF}_{l=1} * S_{0}$. Subsequently, we increase $l$ gradually 
until the full length of the BF is reached ($l = n = 301$).

\begin{figure*}
\vspace{0.cm}\hspace{-0.5cm}\includegraphics[width=10cm]{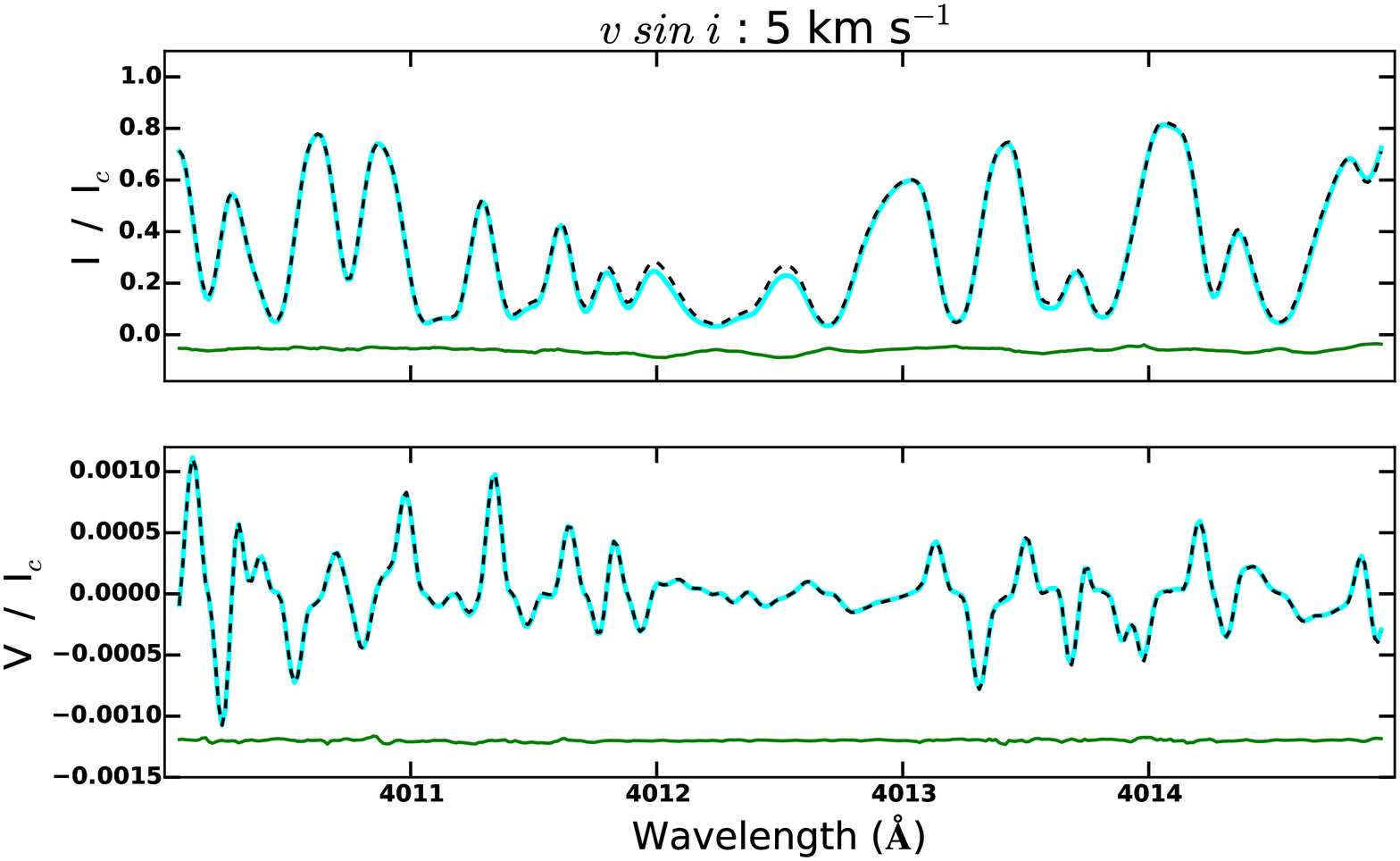}
\includegraphics[width=10cm]{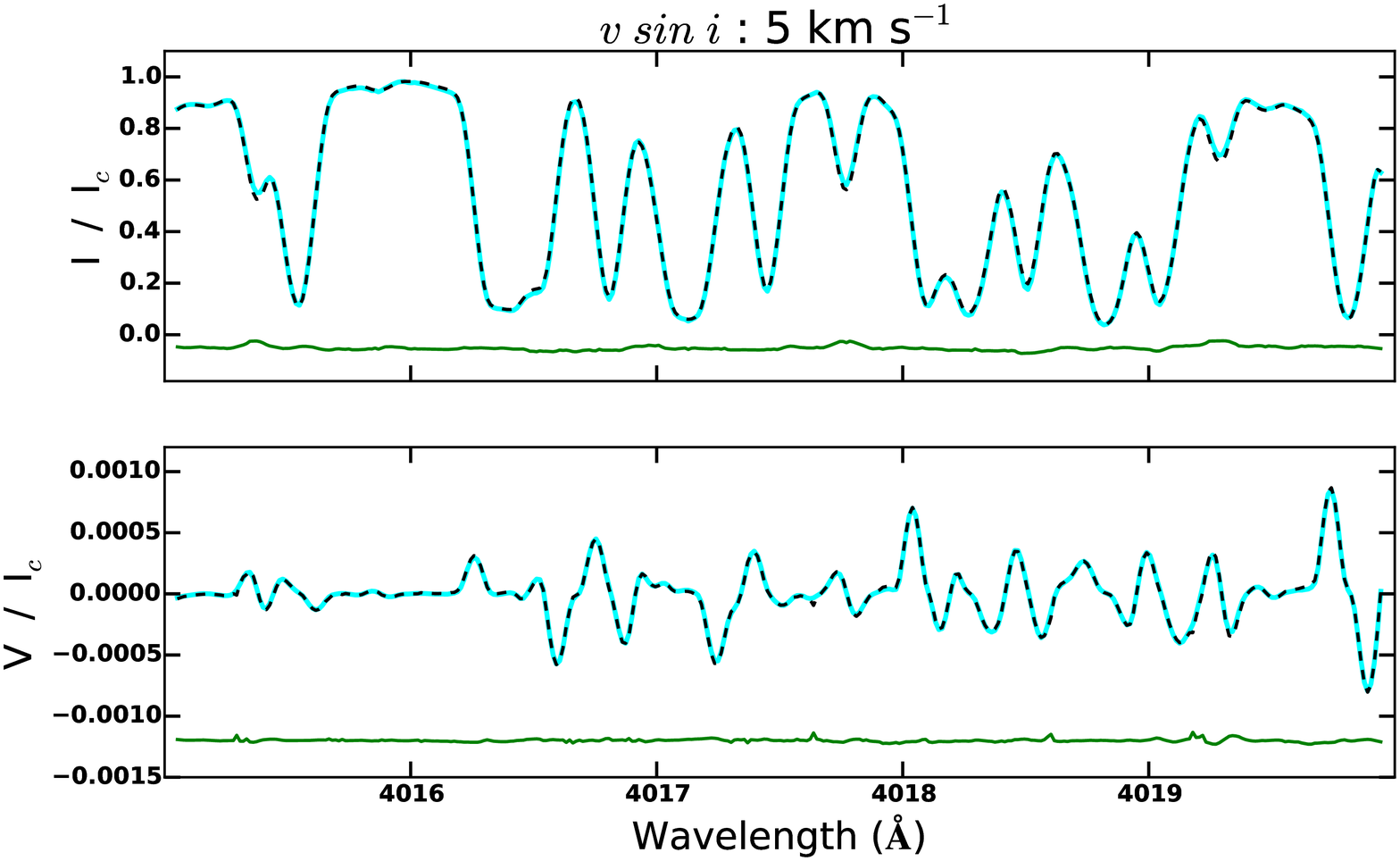}

\vspace{0.5cm}\hspace{-0.5cm}\includegraphics[width=10cm]{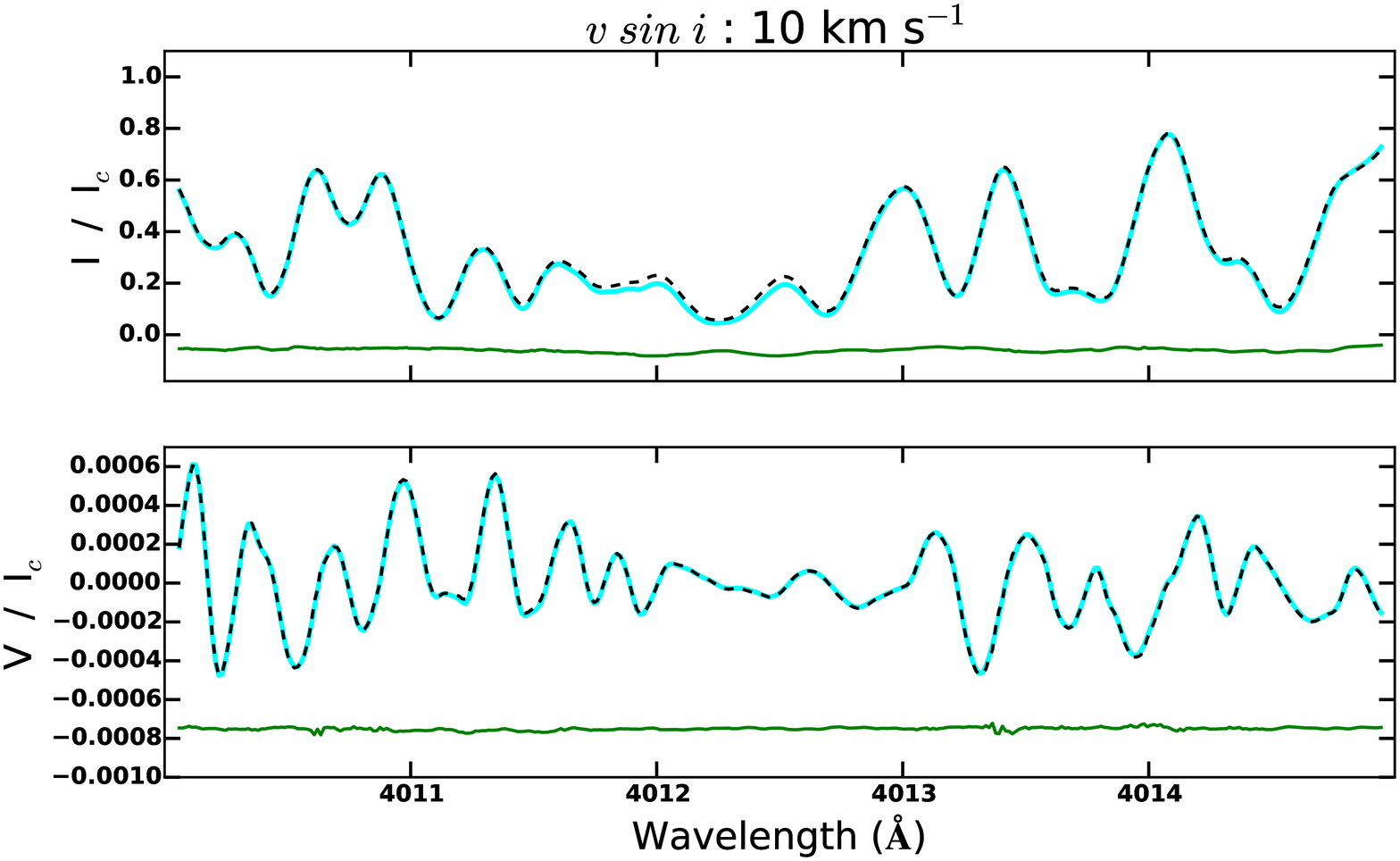}
\includegraphics[width=10cm]{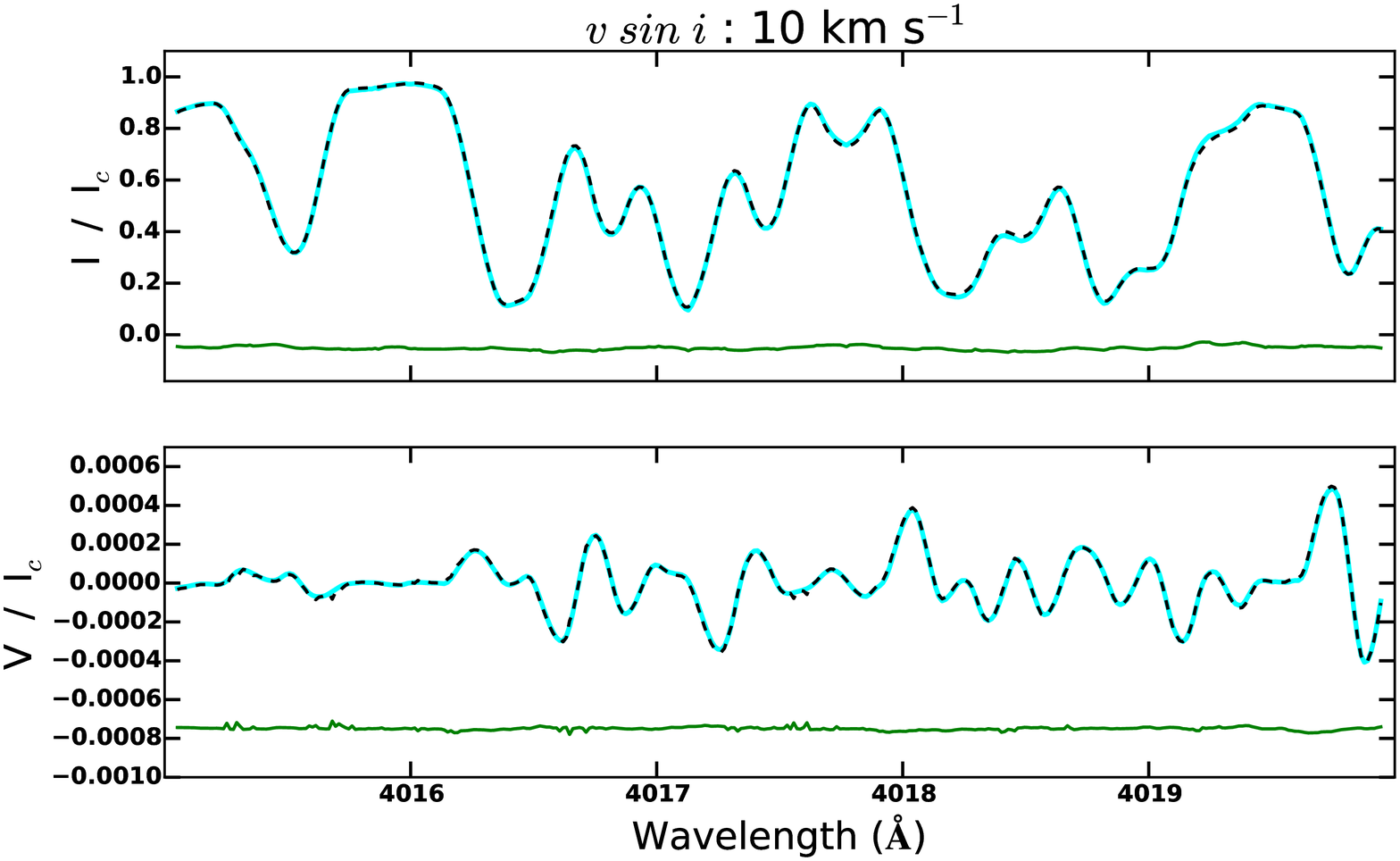}

\vspace{0.5cm}\hspace{-0.5cm}\includegraphics[width=10cm]{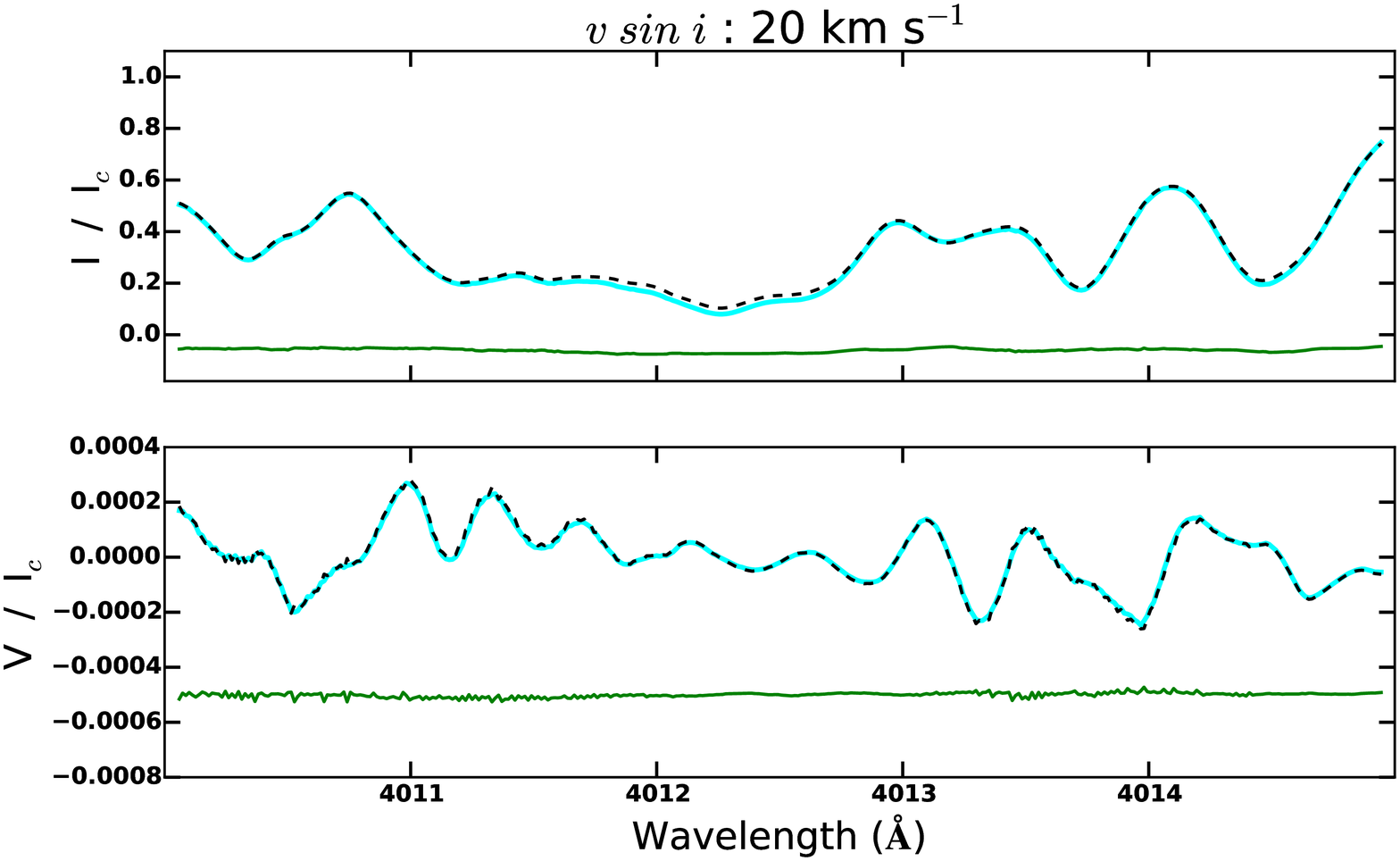}
\includegraphics[width=10cm]{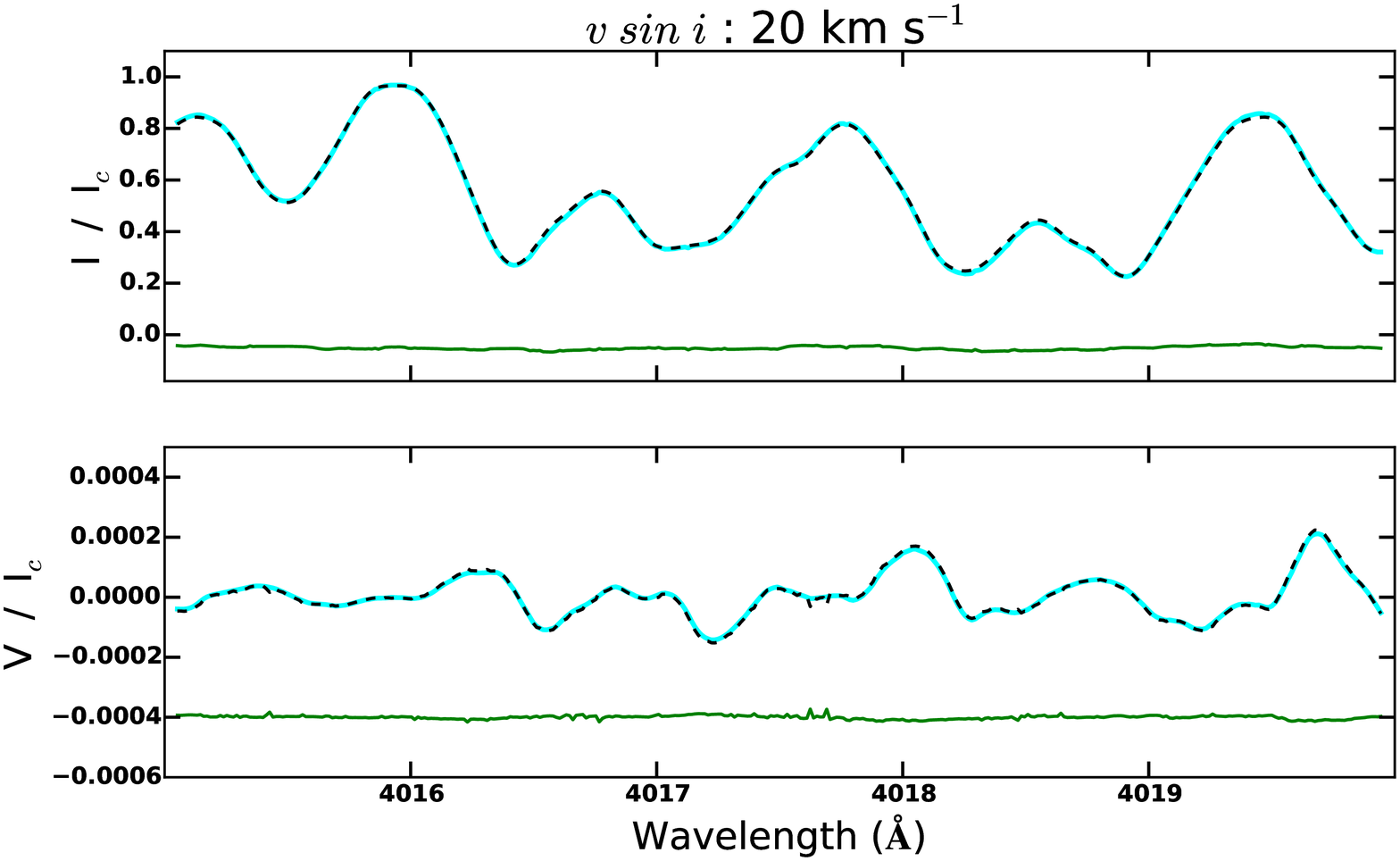}

\caption{Same as in Fig.\,\ref{fig:comp} but considering different 
rotation velocities: from top to bottom 5, 10 and 20\,km\,s$^{-1}$, respectively.}
\label{fig:rota_V}
\end{figure*} 

In Fig.\,\ref{fig:BF} we show the BF using the full length of the spectral 
window (upper left panel), and the one that gives the best fit (upper right 
panel). The accuracy of the fits as a function of the number of eigenvectors 
included while constructing the BF is displayed in the lower panel.  Note 
that the fit of the profiles improves quickly as the number of eigenvectors 
increases, it then enters a plateau phase, subsequently improving once more 
until reaching the optimum fit ($l = 91$), and finally entering a second 
plateau phase. The explanation  for particular behaviour is  twofold. 
From one side, the broadening regions do not contain information about 
the broadening process (e.g. \citet{Rucinski1992}), and from the other side,
the numerical value in these continuum regions is very close to zero but 
is not exactly zero. 
These two properties make that the curve shows this two plateau behaviour.
We have corroborated this fact by adding  noise to the 
profiles, and in this case the fits improve monotonically until the optimum 
fit is reached.
Of course, the fit is better when there is no noise. 
In the noise-free case, the difference between ``solar'' and stellar $V$ 
profiles for the optimum fit is gratifyingly small (MAE of 8.6\,$\times$ 10$^{-6}$), 
and in fact even better than the fits found for the case of $v\,\sin i$ = 0  
listed in Table\,2. It has to be kept in mind that in real observations, 
the latest  eigenvectors are associated with noise. Hence, it is important 
to carefully determine the optimum number of eigenvectors to use in the 
construction of the BF. 
 
We applied the broadening procedure to the Stokes profiles shown in
Fig.\,\ref{fig:comp}  with values of $v\,\sin i$ of 5, 10, and 
20\,km\,s$^{-1}$. It turns out that the adaptive grid is very sensitive to 
the value of $v\,\sin i$. We have therefore modified the parameters of 
the adaptive grid in such a way as to obtain a number of points comparable 
to the one used for the fixed grid (close to 400). The resulting profiles 
are shown in Fig.\,\ref{fig:rota_V} and demonstrate that the broadening 
procedure applied to the ``solar'' profiles is working properly; 
please note that the BF properly reproduces both the broadening due to rotation and the 
broadening due to the distribution of magnetic field strength across the
stellar surface. This will become even clearer in the next section where 
we will consider higher magnetic field strengths.

\section{Multi-line technique}
                                    
We now want to investigate whether it is possible to generalise our results to
spectral ranges of thousands of Angstroms, to strong magnetic fields and to non-zero 
rotational velocities. For this purpose, in what follows we shall employ the PCA-ZDI 
multi-line technique developed by \citet{Semeletal2006}. Details concerning the addition 
properties of individual lines in the context of this technique have been presented in 
\citet{Semeletal2009}, and a detailed description of the procedure used to obtain the 
MZSs and a discussion of some of their properties are given in \citet{Ramirezetal2010}.
Please note that ``Zeeman Doppler Imaging'' is used here in the original sense 
introduced by \citet{SemelSe1989}, not to be confounded with the {\em mapping} of 
magnetic and/or abundance structure as e.g. in \citet{Hussainetal2000}. The PCA-ZDI 
technique is based on a representative database of synthetic Stokes spectra, 
calculated in the present case with the help of {\sc Cossam}; PCA is used to obtain 
the eigenvectors of that database. Subsequently, these eigenvectors serve as detectors 
(in analogy to the line mask used in LSD), in order to obtain the MZSs through a cross 
correlation between the observed spectra and the eigenvectors. In order to prevent 
confusion, let us emphasise that the eigenvectors used in the PCA-ZDI technique have 
nothing to do with the eigenvectors used for the construction of the BFs described
previously. In fact, the eigenvectors used in PCA-ZDI form the basis of the Stokes 
parameters in the synthetic database, such that any Stokes profile in that database 
can be represented by a linear combination of the PCA-ZDI eigenvectors. For more 
details see the three references given above.

With the PCA-ZDI technique it is possible to obtain one MZS per eigenvector, i.e. 
as many as the model spectra contained in the database (a total of 700 for this 
work, see below).  However, the first eigenvector is the most significant one,
exhibiting the most representative line shape pattern of all models included in 
the database. Thus, in order to simplify the interpretation of the results, for our 
tests and analyses we shall employ only the MZS obtained with the first eigenvector.

In addition, given that nowhere in the whole procedure has similarity of the 
individual lines been assumed nor any special regime of the magnetic field strength 
and direction, our technique does not suffer any of the constraints typical for 
LSD. Our only restriction pertains to a minimum ratio of line depth to continuum, 
which for this work we have fixed at $> 0.1$, resulting in a total number of 
individual lines used for the construction of MZSs of $\sim$ 38\,000.

The first magnetic detections with the help  PCA-ZDI in linear and circular 
polarisation were presented a decade ago (\citealt{Semeletal2006},
\citealt{Ramirezetal2006}), but magnetic field quantities like $H_{\rm eff}$ have 
not yet been determined with this technique. In the following we show how 
stellar effective longitudinal magnetic fields can be retrieved by application 
of PCA-ZDI.

\begin{figure}[t]
\hspace{0.2cm}\includegraphics[width=3.3cm]{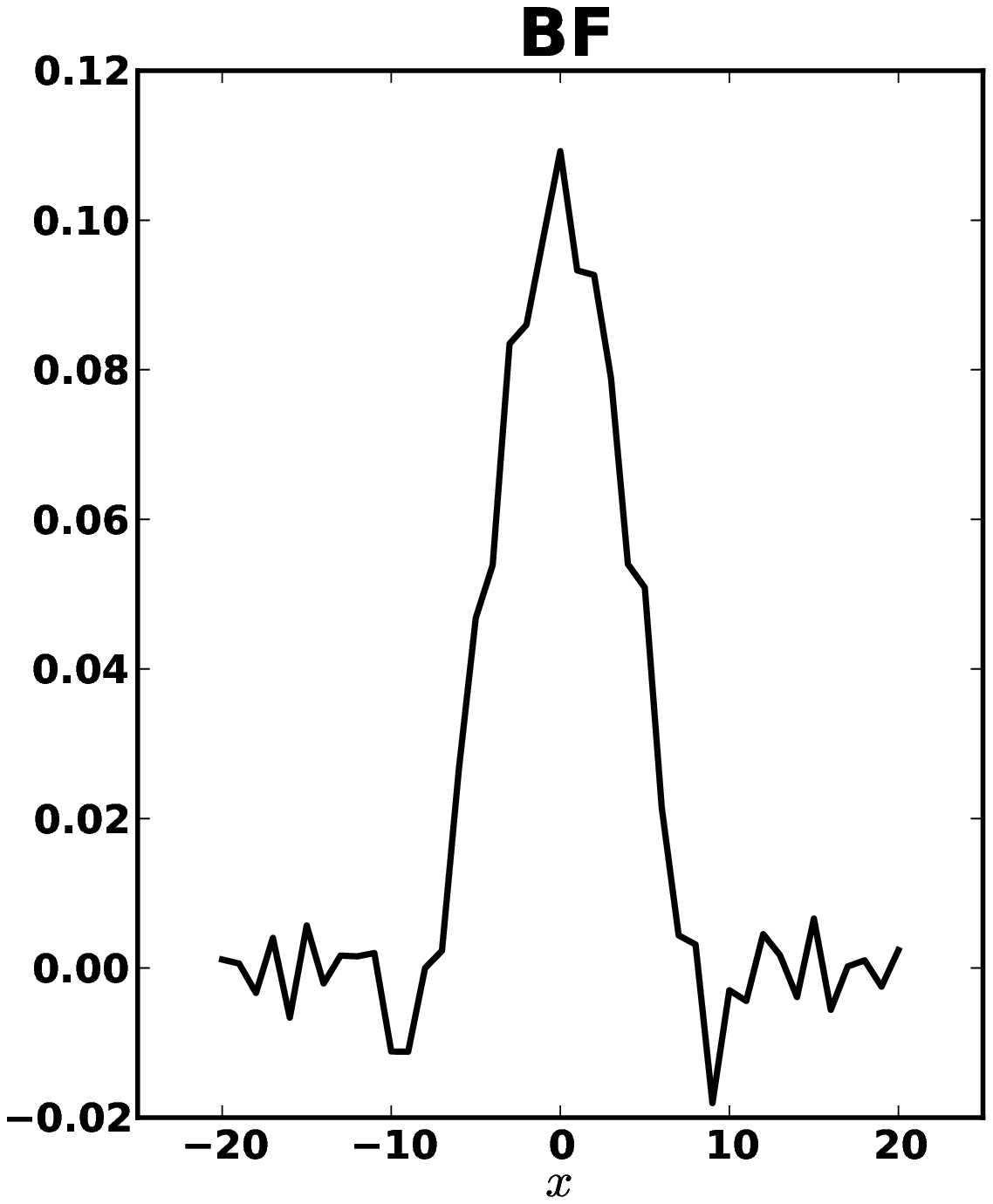}\hspace{0.3cm}\includegraphics[width=5.4cm]{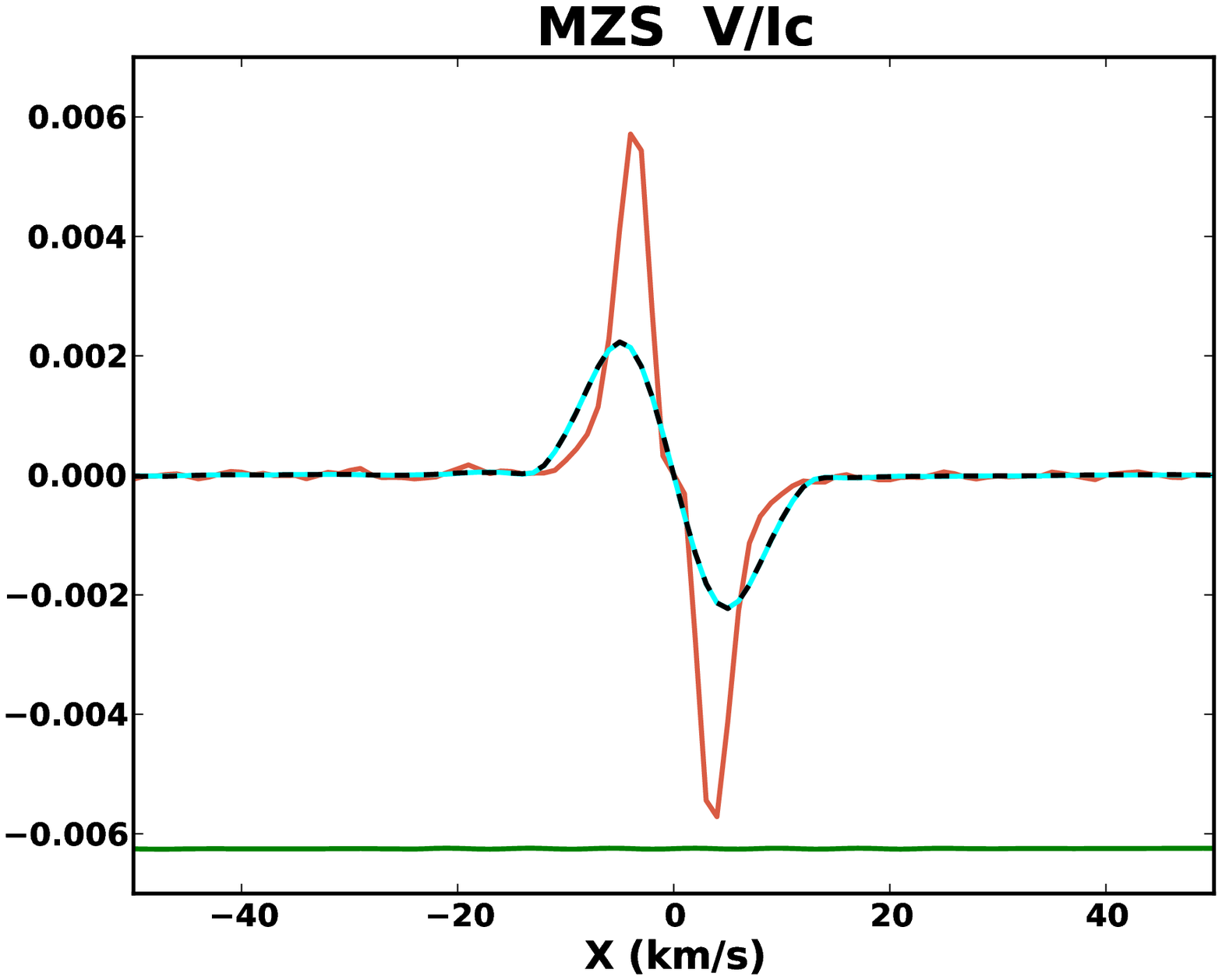}
\caption{Left panel: optimal BF, derived with 41 eigenvectors. Right
panel: MZS profiles. The red line represents the ``solar'' MZS ($v\,\sin i = 0$\,km\,s$^{-1}$), 
the dashed black line the stellar MZS ($v\,\sin i = 10$\,km\,s$^{-1}$), and 
the solid line in cyan the result of the application of the BF to the ``solar'' MZS.  
The green line at the bottom of the panel, shifted for clarity's sake, 
represents the difference between the ``solar'' broadened MZS and the stellar MZS. 
The stellar Stokes profiles were calculated using Model\,1, a dipolar magnetic moment 
of $m=400$\,G and a fixed grid of 440 quadrature points.}
\label{fig:fit_mzs1}
\end{figure}

\begin{figure} [h]
\hspace{0.2cm}\includegraphics[width=3.3cm]{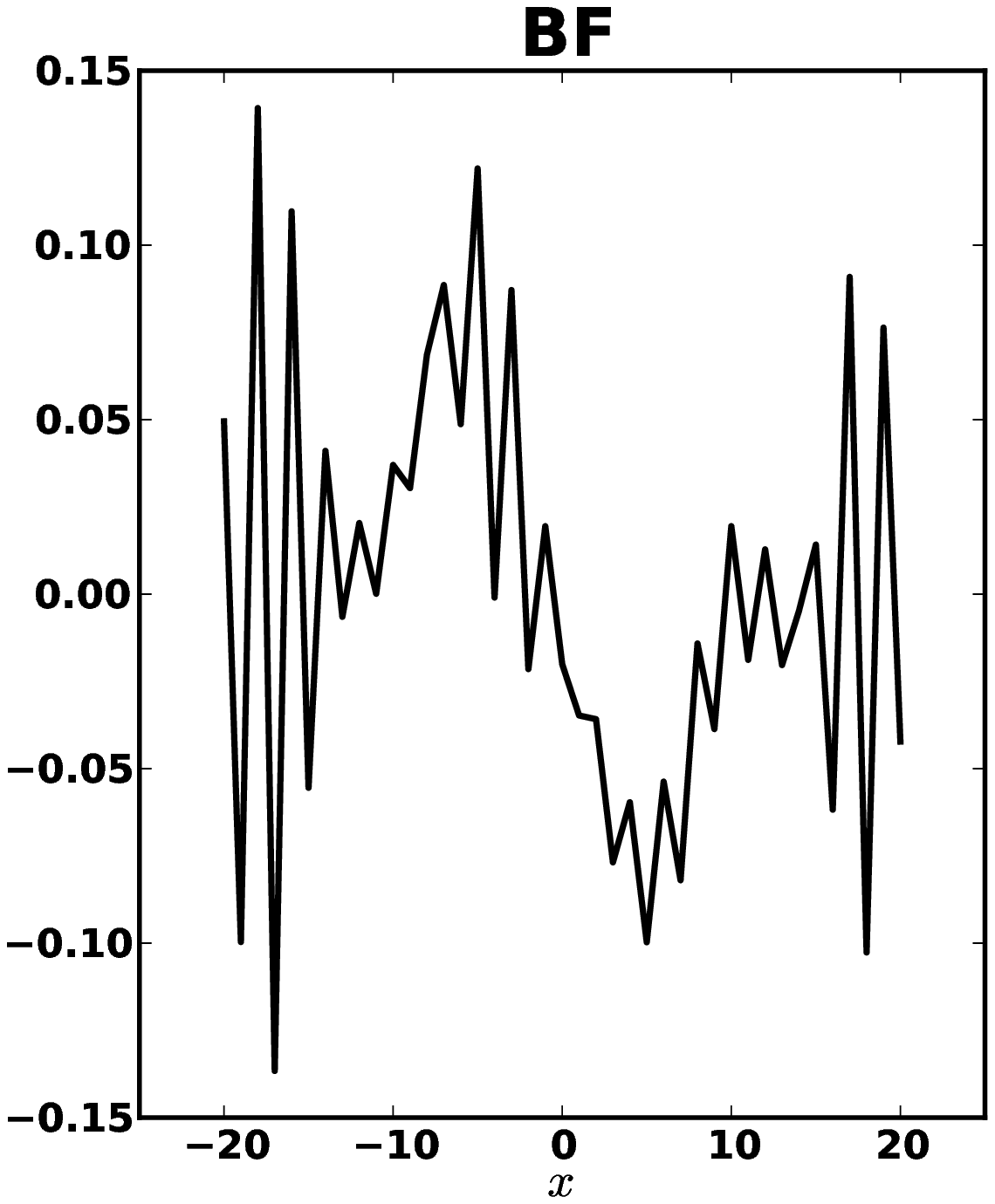}\hspace{0.3cm}\includegraphics[width=5.4cm]{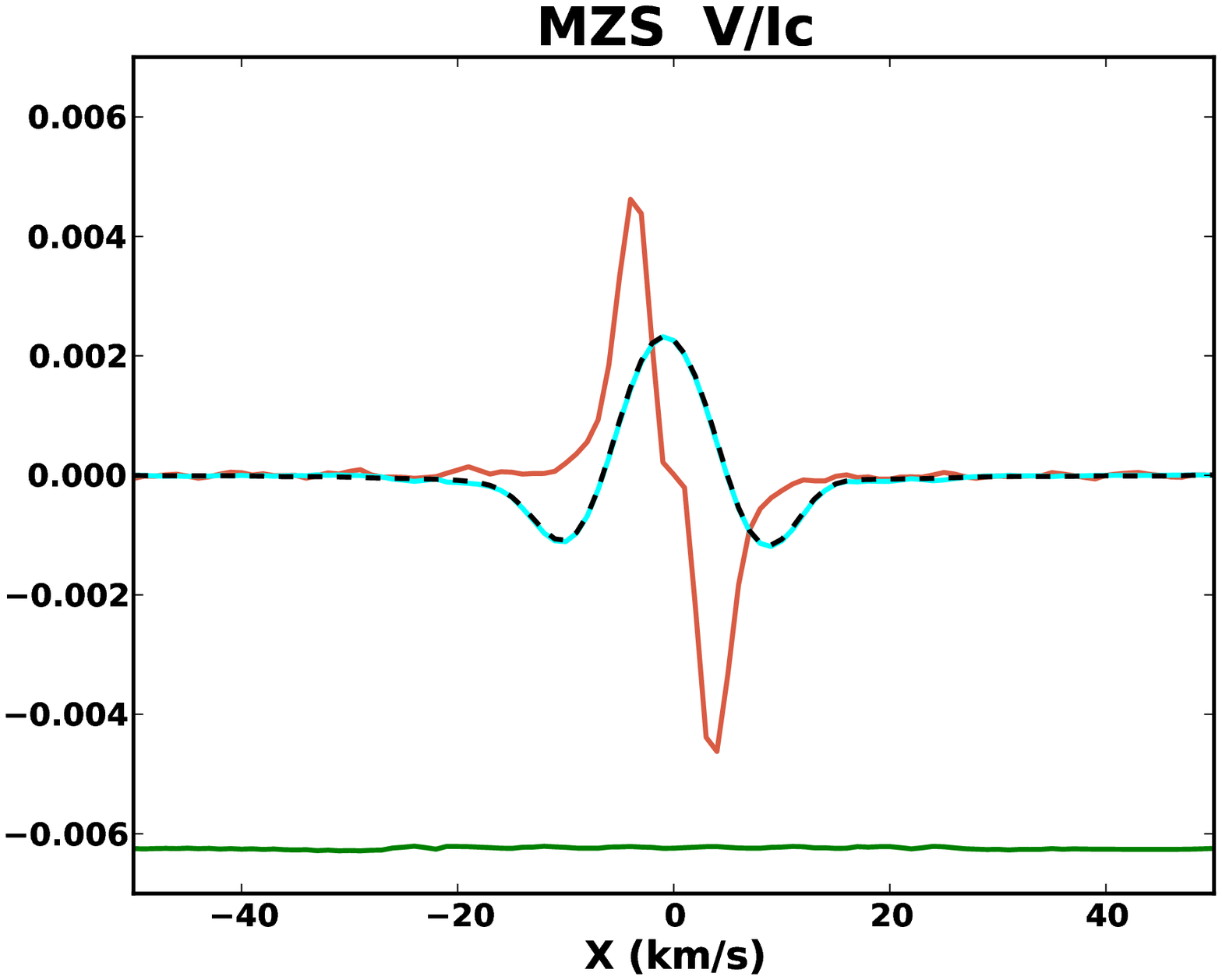}
\caption{Same as Fig. \ref{fig:fit_mzs1} but referring to Model\,2
 (see Table \ref{tab:magmod}).}
\label{fig:fit_mzs2}
\end{figure}

\subsection{MZS using PCA-ZDI}

We have considered a wide spectral range from 350 to 1000\,nm in steps of 
1\,km\,s$^{-1}$ (yielding a total of $\sim$ 315\,000 wavelength points). Adopting
the same atmospheric model as before, we built a database of Stokes profiles
for the ``solar'' case, with the magnetic field varying between 0 and 350\,G in
steps of 0.5\,G. From this database we obtained the eigenvectors,  but we 
then use only the first eigenvector to  produce a total of 700 ``solar'' MZSs. 
Our goal is to retrieve  $H_{\rm eff}$ -- a ``stellar'' (integrated) quantity -- 
by converting ``solar'' MZSs to stellar MZSs, similarly to what we did in the 
preceding section.

For this purpose, let us consider the stellar MZS obtained from Model\,1 of
Table\,\ref{tab:magmod}, $v\,\sin i = 10$\,km\,s$^{-1}$ and a magnetic moment of
$m = 400$\,G. The effective field for this configuration at phase zero is 256\,G. 
We then search in the ``solar'' space of MZSs for a model with field strength
equal to the value of $H_{\rm eff}$. Now we are confronted with the same situation
as before, i.e. we have to apply the proper broadening to the ``solar'' MZS to
reproduce the stellar MZS but with the difference that instead of working with
spectral lines, viz. Eq.\,\ref{ecBF}, we work with the MZSs:
\begin{equation}
 MZS_{\rm rot}(X) = BF(X) * MZS_{\rm p}(X),
\label{BFmzs}
\end{equation}
where $MZS_{\rm rot}$ stands for the stellar MZS, $MZS_{\rm p}$ for the ``solar''
MZS and $X$ indicates that we are working in Doppler space.

The broadening function is found as explained in the preceding section (\ref{sec2.3}). 
In the left panel of Fig.\,\ref{fig:fit_mzs1} we show the optimum BF and in the right
panel the ``solar'' MZS (red line), the stellar MZS (dashed black line), and the
``solar'' broadened MZS (full line in cyan). The difference between the last two
MZSs is shown at the bottom of the panel (full line in green).

\begin{figure}[t]
\vspace{-0.2cm}\hspace{0.4cm}\includegraphics[width=8.5cm]{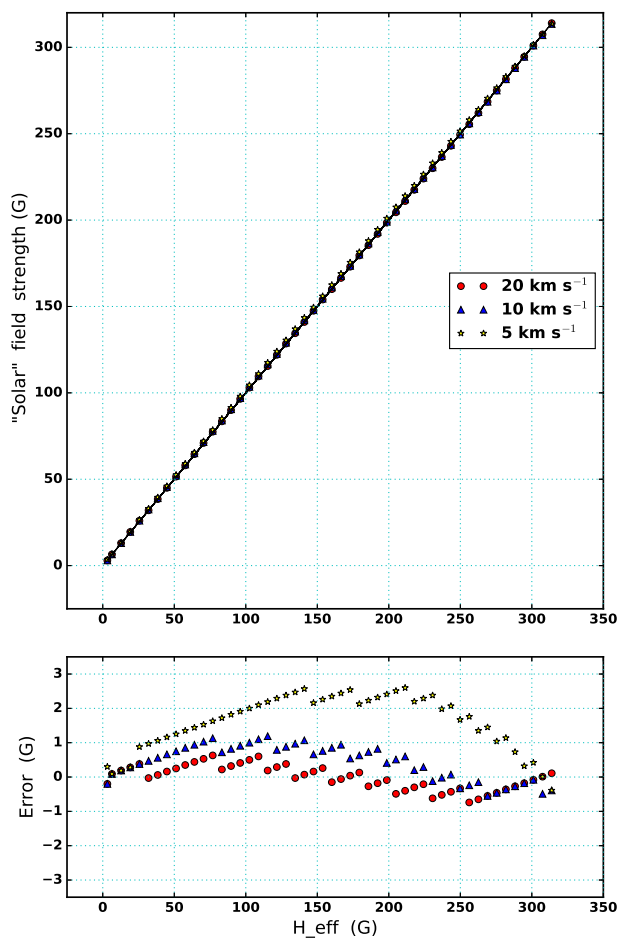}
\caption{Upper panel: Inversion of MZSs using the fixed grid with 
 a centred dipole pointing towards the observer (Model\,1 of Table \ref{tab:magmod}).  
 The various values of $v\,\sin i$ are shown with different symbols and colours.
 Lower panel: Inversion errors.}
\label{fig:inv_l}
\end{figure}

We now want to demonstrate that a BF established this way can be applied irrespective
of the strength of the effective longitudinal field $H_{\rm eff}$. To this purpose, we
simply apply the BF to all ``solar'' MZSs and we look whether the $H_{\rm eff}$ values
for a set of stellar MZSs with different magnetic field strengths can correctly be
recovered. Please note that the inversions concern only circular Stokes profiles, they
do not include the MZSs in intensity. Varying the magnetic moment (m = [5, 10, 20, 30
... 490\,G]), we produce 50 stellar MZSs, carrying out this procedure for the three
values of $v\,\sin i$ previously used: 5, 10 and 20\,km\,s$^{-1}$. It has to be
emphasised that for each value of $v\,\sin i$ it is mandatory to find the appropriate
BF. The results displayed in Fig.\,\ref{fig:inv_l} clearly show that our approach leads 
to the correct retrieval of the stellar $H_{\rm eff}$. 

Nevertheless, it is important to realise that the BF also depends on the orientations 
of the principal axes of the system -- combinations of the 3 Eulerian angles $\alpha$, 
$\beta$, $\gamma$ and of the inclination angle $i$ between rotation axis and the LOS. To 
illustrate this dependence, we take the example shown in Fig.\,\ref{fig:fit_mzs1} and we 
recalculate it with the only difference that the angles of the system change from those 
of Model\,1 to those of Model\,2. For this new configuration, the longitudinal field 
value drops to $H_{\rm eff} = 212$\,G, but much more importantly, the shape of the MZSs
and the BF change drastically (see Fig.\,\ref{fig:fit_mzs2}).

We have established that for the recovery of the effective longitudinal field
$H_{\rm eff}$, one cannot use some arbitrary BFs. In other words, not only for each
rotational velocity, but also for each orientation of the principal axes of the system 
there exists a particular associated BF that allows the proper recovery of $H_{\rm eff}$. 
To illustrate this fact, we repeat the field inversions of the 50 stellar MZSs 
calculated with Model\,1, but this time to broaden the ``solar'' MZSs we used the BF 
of Model\,2  (left panel of Fig.\,\ref{fig:fit_mzs2}). 
From the results (not shown) it becomes abundantly 
clear that the field inversions fail completely. In practice, when dealing with 
observational data, it is not possible to know a priori the orientation of the 
principal axes. We thus have to devise a method that generalises the field inversion 
procedure to any arbitrary orientation.

\subsection{Magnetic field inversions for arbitrary orientations}

\begin{figure*}[ht]
\hspace{1.5cm}\includegraphics[width=15cm]{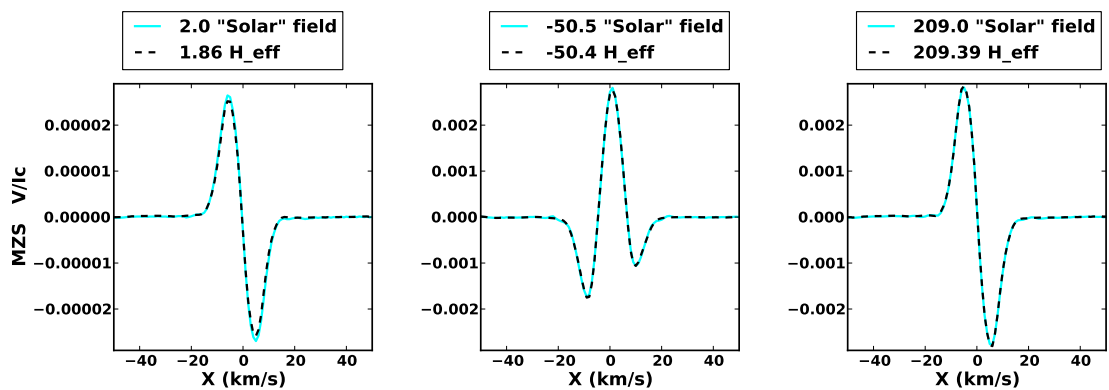}
\caption{Examples of 3 stellar MZSs (dashed black lines) and their 
   respective inversions (solid lines in light colour). The values of $H_{\rm eff}$ are 
   well retrieved in all cases as shown in the legends of each panel. Both the ``solar'' 
   field strength and $H_{\rm eff}$ are given in Gauss.}
\label{fig:fit_mzs_mvariable}
\end{figure*}

We now want to demonstrate that it is possible to correctly recover the effective
longitudinal field from MZSs for arbitrary combinations of the angles $\alpha$, 
$\beta$, $\gamma$ and of the inclination $i$. Considering $v\,\sin i = 10$\,km\,s$^{-1}$, 
adopting a magnetic moment of $m = 400$\,G (not implying any loss of generality) and 
randomly varying the four angles, we calculated 600 different stellar MZSs at phase 
zero. For each combination of the angles, we obtained a particular stellar MZS with 
an attached value of $H_{\rm eff}$ and an associated BF. Each of these 600 BFs 
were obtained following the same process that we employed in Figs.\,\ref{fig:fit_mzs1} 
and \ref{fig:fit_mzs2}, i.e. by solving Eq.\,(\ref{BFmzs}) based on the method 
described in section\,\ref{sec2.3}. All the BFs were constructed consistently with 
the same number (41) of eigenvectors. Now let BF$_{n}$ denote the broadening function 
associated with the n-th combination of angles.  This BF$_{n}$ was directly 
applied to the 700 ``solar'' MZSs -- each one having a different value of field 
strength from 0 to 350\,G in steps of 0.5\,G.   Repeating the process for all the 
BFs results in a total of $600 \times 700$ ``solar'' broadened MZSs, all of which 
are characterised by a particular combination of angles and magnetic field strength.

\begin{figure}[h]
\includegraphics[width=9cm]{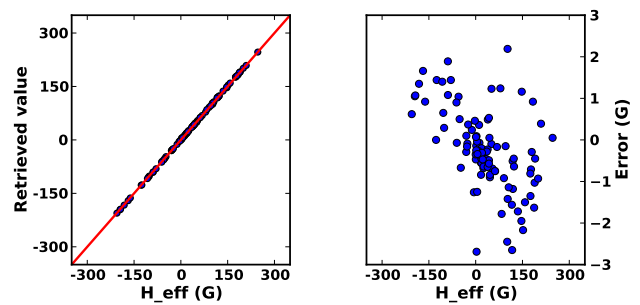}
\caption{Recovery of $H_{\rm eff}$ from stellar MZSs (at left) and
uncertainties (at right), considering $\alpha$, $\beta$, $\gamma$, $i$ and the magnetic 
moment $m$ as variable parameters. The MZSs were calculated using an adaptive grid 
($\sim$80 points).}
\label{fig:mzs_mvariable}
\end{figure}

Considering the same rotation value of $v\,\sin i = 10$\,km\,s$^{-1}$\,, we 
then calculated a set of 100 stellar MZSs with random variations of the four angles 
$\alpha$, $\beta$, $\gamma$, $i$ and of the dipolar magnetic moment -- between 0 
and 490\,G. We inverted these stellar MZSs in the database made up of the broadened 
``solar'' MZSs. Fig.\,\ref{fig:fit_mzs_mvariable} shows detailed fits of 3 among 
the 100 stellar MZSs, suggesting that our proposed technique is able to fairly 
accurately recover the effective longitudinal field of a star. The results for 
the whole bulk of 100 stellar MZSs is displayed in Fig.\,\ref{fig:mzs_mvariable} 
and definitely validates the excellent performance of the method, reflected in 
gratifyingly small errors. 

These encouraging results were obtained, implicitly, only for the special case when 
the dipole position is known a priori (in our latest example at the centre of the 
star). More generally, one has however to consider all the parameters of the magnetic 
model to be unknown, a problem which we shall address in the following.

\subsection{Retrieving $H_{\rm eff}$ in the general case}

We performed a similar test to the one presented in the preceding section but now 
including as unknowns the position and direction of the dipole. 
We calculated 7500 stellar MZSs, varying randomly all the parameters in the 
magnetic model ($\alpha$, $\beta$, $\gamma$, $i$, $x_2$, $x_3$). The position of the 
dipole was constrained to less than 0.3\,$R_{\star}$. The angles ($\alpha$, $\beta$, 
$\gamma$) were allowed to vary between $-180\deg$ and $+180\deg$, and  the inclination 
$i$ between $0\deg$ and $+180\deg$. Please note that with these limits to the range
of the Eulerian angles, there is a certain redundancy in the resulting models, so
that we can fix the rotational phase arbitrarily at zero. Finally, we arbitrarily
adopted a magnetic moment of 400\,G as before and we also assumed a value of 
$v\,\sin i = 10$\,km\,s$^{-1}$.

\begin{figure}[h]
\vspace{-0.35cm}\hspace{0.15cm}
\includegraphics[width=8.2cm]{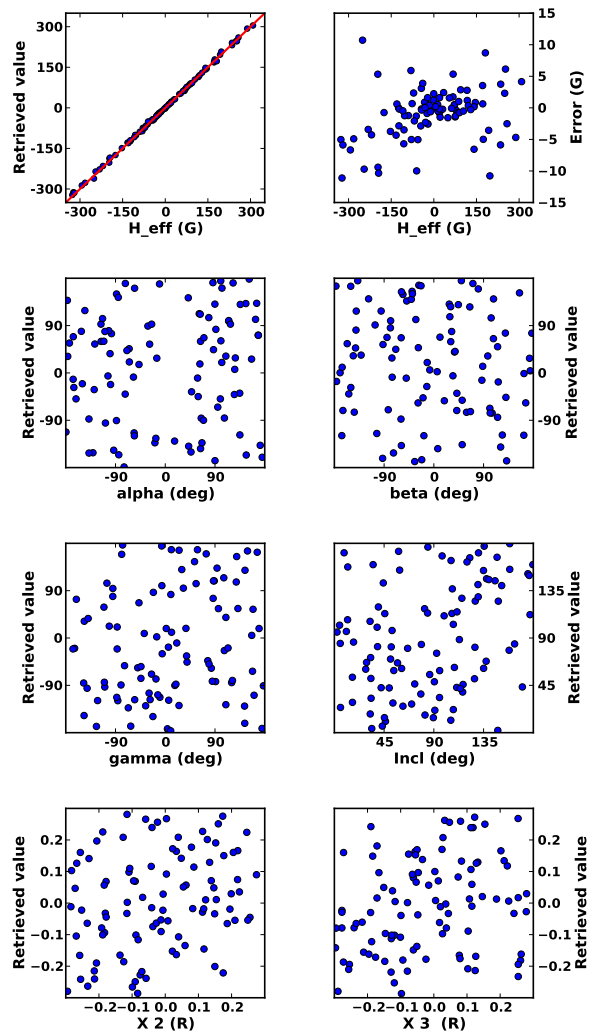}
\vspace{-0.3cm} \caption{Inversion -- in the general case -- of stellar MZSs, using a fixed 
grid (80 points) and considering all the parameters in the magnetic model ($\alpha$,
$\beta$, $\gamma$, $i$, $x_2$, $x_3$, $m$) as unknowns. In each panel the x-axis
gives the values of the input parameters to the stellar MZS. The y-axis of the
upper right panel gives the inversion errors for the $H_{\rm eff}$, for the rest of 
the panels it gives the parameter values retrieved by the inversions.}
\label{fig:mzs_mvariable2}
\end{figure}

For each of these MZS we repeated the procedure described in the last section, i,e. 
for any given combination of magnetic parameters we obtained an associated MZS with 
a particular value of $H_{\rm eff}$ and an attached BF. Considering all the calculated 
magnetic models we obtained a total of 7500 BFs which we applied to broaden the 
``solar'' MZSs, reaching a total of $7500 \times 700$ ``solar'' broadened MZSs. 
Again, each of these ``solar'' MZSs is distinguished by a particular combination 
of magnetic geometry and field strength. Then, to test the inversions, we 
calculated a set of 100 stellar MZSs adopting $v\,\sin i = 10$\,km\,s$^{-1}$ and 
random values of the four angles, of the dipole moment (between 0 and 690\,G) and 
of the dipole position ($< 0.3 \, R_{\star}$). We show the results of the inversions 
in Fig.\,\ref{fig:mzs_mvariable2}.

These plots reveal that none of the geometric model parameters ($\alpha$, $\beta$,
$\gamma$, $i$, $x_2$, $x_3$) can be correctly retrieved. This is not surprising 
since for a given value of $H_{\rm eff}$ a multiplicity of combinations of angles, 
magnetic moments and dipole positions can produce the same stellar longitudinal 
field. Nevertheless the value of $H_{\rm eff}$ is in general well retrieved. The 
inversion errors for $H_{\rm eff}$ are larger than in the case of a fixed dipole 
position.  Besides, in our tests we have considered as fixed the atmospheric model
 ($T_{\rm eff}$, $log\,g$, $[M/H]$, $v_{\rm turb}$, $\xi$)
and for this reason we have not included Stokes $I$ in the inversions; 
when inverting simultaneously the Stokes $I$ and $V$ parameters, the case of real data,  the inversion incertitudes
will increase. However, please note that a way to reduce the inversions errors is  to increase  
the number of BFs considered (7500 in this test). We thus may conclude that it is possible with our
method to correctly measure stellar longitudinal magnetic fields.

\subsection{Spatial grid resolution with PCA-ZDI profiles}

Throughout this work we have used both adaptive and fixed grids at different 
spatial resolutions. Now we want to show that the database used in the last test,
based on 80 spatial quadrature points, can be used to invert MZSs calculated at 
much higher spatial grid resolution (up to 1000 points and more). For this purpose 
we have considered two arbitrary magnetic models whose parameters are listed in 
Table \ref{tab:grids}. Taking an adaptive grid with 80 quadrature points and a 
magnetic moment of 39.26\,G, for Model\,1 one obtains $H_{\rm eff} = -24.32$\,G
at phase zero. Similarly, adopting a value of $m = 395.78$\,G, we find 
$H_{\rm eff} = 228.90$\,G at phase 0.526 for Model\,2.

\begin{figure}[b]
\hspace{-0.25cm}\includegraphics[width=9cm]{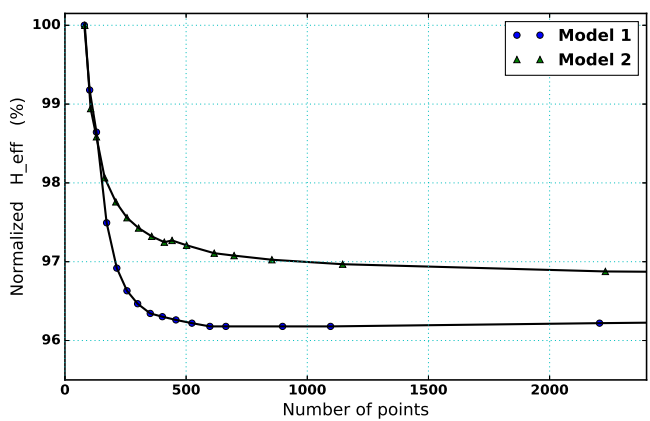}
\caption{Percentual variation of $H_{\rm eff}$ as a function of the 
   number of quadrature points considered in the adaptive grid. The  $H_{\rm eff}$ values 
   are normalised to the $H_{\rm eff}$ value obtained with a grid of 80 points.}
\label{heff_grid}
\end{figure}

As mentioned right at the beginning, $H_{\rm eff}$ is a quantity that depends on 
the limb darkening, being thus potentially affected by the number and the spatial 
distribution of the quadrature points. We have inspected this problem with the
help of the two magnetic models of Table \ref{tab:grids}, varying the grid 
resolution. In Fig. \ref{heff_grid} we plot $H_{\rm eff}$ as a function of the 
number of grid points; the values of $H_{\rm eff}$ are given relative to the value 
of $H_{\rm eff}$ calculated with an 80 point grid. We see a modest but significant 
change in $H_{\rm eff}$ when the number of grid points is increased, reaching a 
difference close to 4\% and 3\% respectively for Models\,1 and 2 at $\sim$\,500 
grid points. Further increasing the number of points gives only very small
variations in $H_{\rm eff}$. For Model\,1 $H_{\rm eff}$ remains essentially constant 
when the number of grid points is in excess of 600, whereas for Model\,2 
the normalised longitudinal field passes from 97.2\% to 96.9\% between 502
and 1096 grid points. We therefore consider a grid resolution of about 1000 
points sufficient for highly accurate $H_{\rm eff}$ determinations in both models.

\begin{table}[t]
\caption{Magnetic models used to test the spatial grid resolution.}
\label{tab:grids}
 \begin{tabular}{|c|cccccc|}
  \hline
       &  i ($^{\circ}$) & $\alpha$ ($^{\circ}$) & $\beta$ ($^{\circ}$) & $\gamma$ ($^{\circ}$) 
       & $x_2$ ($R_{\star}$) & $x_3$ ($R_{\star}$) \\
  \hline
 Model 1 & 124.7 &  107.4  & 16.5  & -20.9 &  0.22 & -0.11 \\  
 Model 2 &  26.1 &  -56.3  & 25.0  &  63.6 &  0.18 &  0.20 \\  
 \hline
 \end{tabular}
 \end{table}

\begin{figure}[t]
\hspace{0.1cm}\includegraphics[width=9cm]{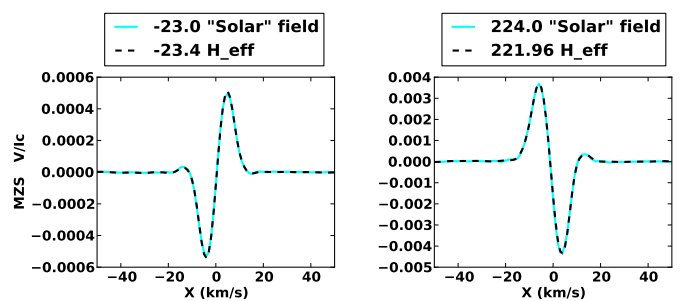}
\caption{Examples of 2 stellar MZSs constructed at very high 
    grid resolution ($>$\,1000 points) using adaptive grids. The dashed black lines 
    represent the stellar MZSs while the solid lines in cyan colour are their 
    respective inversions. The retrieved values of the $H_{\rm eff}$ (shown at the
    top of the panels) clearly indicate that the PCA-ZDI profiles are quite insensitive 
    to the grid resolution.}
\label{hr_stift}
\end{figure}

\begin{figure}[]
\vspace{0.3cm}\hspace{0.1cm}\includegraphics[width=9cm]{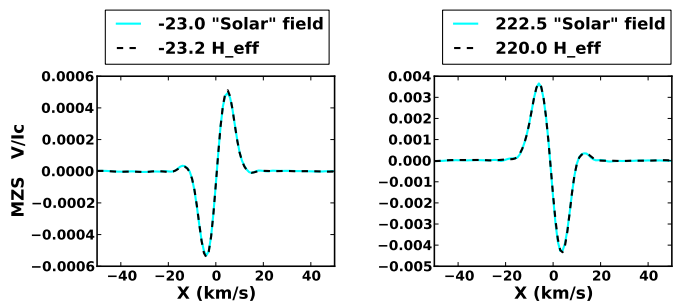}
\caption{Same as Fig. \ref{hr_stift} but considering a fixed
    grid of 1100 points in the construction of the stellar MZSs.}
\label{hr_lands}
\end{figure}

We thus established the stellar MZSs for Models\,1 and 2 at high resolutions (1096 
and 1146 points respectively), employing an adaptive grid. We then inverted these 
two MZSs in the ``solar'' broadened database used for the test presented in the 
preceding section. Please note that the BFs applied to broaden the ``solar'' MZSs of 
that database were obtained from stellar MZSs established with a fixed grid of 
80 points. We then performed the same inversions again, but with only difference
that now we used a fixed grid of 1100 points to establish the two stellar MZSs. 
The results shown in Figs. \ref{hr_stift} and \ref{hr_lands} clearly reveal that 
we can properly retrieve $H_{\rm eff}$ from the stellar MZSs constructed at high 
grid resolution.

Let us explain the reasons for this gratifying result. After inspection of the $V$ 
profiles at different grid resolutions, we found that in general differences in the 
integrated stellar Stokes profiles are small; only for very few lines the spatial 
grid resolution can have some influence. For example, comparing the adaptive grids 
at high and low resolution, 1046 and 80 points respectively, for Model\,1 we found 
a MAE of the $V$ profiles of $4.5\, \times 10^{-5}$, whereas the maximum difference 
in the line profiles is 2.3\%. Accordingly, when constructing the MZSs through the 
addition of thousands of lines, the differences in the Stokes profiles will cancel 
out statistically, permitting the correct retrieval of $H_{\rm eff}$. One would 
however expect the resolution of the spatial grid to have more important repercussions
on the synthesis of Stokes profiles in the presence of very strong magnetic fields, 
fast stellar rotation or strong stellar pulsations (e.g. \citealt{Fensl1995}), but 
this is a case outside the scope of the present study.

To conclude, the results of Figs. \ref{hr_stift}  and \ref{hr_lands} are very 
encouraging since the most time consuming part of our method lies in the computation 
of thousands of stellar MZSs that serve to obtain the BFs that we use to broaden the 
``solar'' MZSs. The fact that for the calculation of the stellar MZSs we can employ a 
relatively low spatial grid resolution without compromising the results makes our
approach less expensive than might have been feared.

\section{PCA-ZDI and (magnetic) broadening}

\begin{figure*}[ht]
\hspace{0.5cm}\includegraphics[width=17cm]{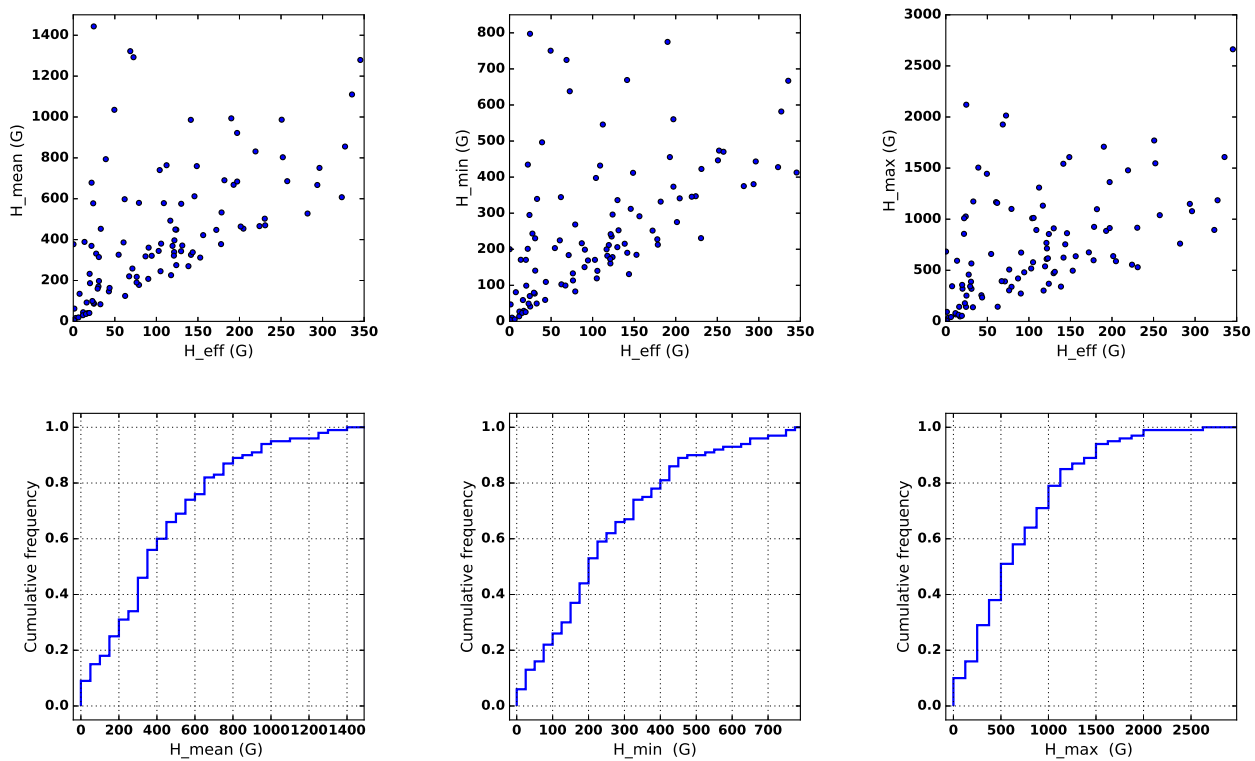}
\caption{  Upper panels, from left to right: mean, minimum 
and maximum surface values of the field modulus as a function of the absolute value of
$H_{\rm eff}$ for the set of 100 MZSs used in the inversion test of Fig. 
\ref{fig:mzs_mvariable2}.  Lower panels: Respective cumulative histograms.} 
\label{fig:hsup}
\end{figure*}

 Given that we obtained the stellar spectra through the integration 
of the local Stokes profiles is that the surface distribution of the local 
magnetic fields is known.

Let us for example have a look at the 100 MZSs employed for the inversion test shown in 
Fig.\,\ref{fig:mzs_mvariable2}. Recall that for the establishment of this set of 
MZSs all the parameters of the magnetic model were randomly varied. In the upper 
panels of  Fig.\,\ref{fig:hsup} we plot the mean, minimum and maximum values of 
the field modulus as a function of $|H_{\rm eff}|$ 
for the full set of MZSs. The lower panels display their respective cumulative distributions. 
Not surprisingly, small absolute values of $H_{\rm eff}$ can be accompanied by 
large field moduli which means that $H_{\rm eff}$ cannot be taken as an indicator 
for the validity of the weak-field approximation. In our sample for example we 
encounter a value of $H_{\rm eff} = 24.6$\,G, associated with mean, minimum and 
maximum values of 1443, 797, and 2120\,G respectively. 
Unlike other multi-line techniques, the PCA-ZDI 
technique allows us to correctly recover $H_{\rm eff}$ irrespective of
the distribution and intensities of the local magnetic fields  
(see upper panels of Fig.\,\ref{fig:mzs_mvariable2}).

To summarise, our approach does not require any of the following assumptions: 
\vspace*{-2mm}
\begin{enumerate}
\item That the local residual Stokes $I$ intensities are small (see e.g. discussion 
      by \citealt{Sennhauseretal2009} and by \citealt{KochukhovMaPi2010}),
\item That the magnetic broadening is largely inferior to the non-magnetic broadening 
      (thermal, rotational, micro/macro turbulence),
\item That the Zeeman splitting results in a normal triplet for all lines.
\end{enumerate}
In other words, with the PCA-ZDI technique we can properly deal with blended 
lines, take into account the anomalous Zeeman effect or negative Land{\'e} factors 
for individual lines, and work in the presence of strong magnetic fields such that 
the spectral lines are in the regime of Zeeman saturation, i.e. the amplitudes of 
the $V$ profiles do not grow anymore linearly with the intensity of the magnetic 
field. This considerably widens the application range of our method compared to
other multi-line techniques like for example LSD.

In addition, we have verified that the weak field approximation 
can not be applied to retrieve $H_{\rm eff}$ from the MZSs. 
We find that the shape of the derivative of the Stokes $I$ MZS does 
not correspond to the shape of the Stokes $V$ MZS, indicating that PCA-ZDI is 
incompatible with the WFA. The reason for this, is simply that our technique 
does not take into account the requiered asummptions of the WFA to be valid.

Finally, we would like to emphasise that the fits to the MZSs 
(Figs.\,\ref{fig:fit_mzs_mvariable},\,\ref{hr_stift},\,\ref{hr_lands}) 
clearly indicate that magnetic and rotational broadening are simultaneously
well been taken into account by the BF method.

\section{Conclusions}

Spectropolarimetry constitutes the optimum observational technique for the study 
of solar and stellar magnetism. These data are best analysed with the help of 
codes that implement spectral line synthesis in the presence of magnetic fields, 
solving the coupled equations of polarised radiative transfer.  It is highly 
desirable that any multi-line approach to the determination of global magnetic 
quantities such as the effective longitudinal field $H_{\rm eff}$ or the mean 
magnetic field modulus $H_{\rm s}$ should incorporate full opacity sampling and 
a correct treatment of polarised radiative transfer when calculating local Stokes 
$IQUV$ profiles. This is exactly what we have done with the help of a database 
of Stokes profiles established with the help of the polarised line synthesis code 
{\sc Cossam} and MZSs obtained by means of the PCA-ZDI technique. 

Efforts to derive $H_{\rm eff}$ from MZSs obtained with PCA-ZDI date back to 
\citet{Semeletal2009} who showed that the centre of gravity method can be applied 
to the MZSs for an estimate of magnetic field strengths.  Additionally, using 
all four Stokes parameters, \citet{Ramirezetal2010} demonstrated that MZSs correctly 
encode the information on the magnetic field, meaning that both strength and 
orientation of the magnetic field can successfully be recovered for field strengths 
of up to 10\,kG. In these two papers, the results were obtained using ``solar'' 
MZSs, i.e. any effect of rotation was neglected and no inhomogeneous spatial 
distribution of magnetic strengths over the surface was assumed.

The present study introduces a novel approach that can be applied to the analysis 
of spectropolarimetric data. The principal idea behind this method consists in the use 
of broadened ``solar'' MZSs to infer the effective longitudinal field $H_{\rm eff}$. We 
have shown that the BFs are an effective tool to properly reproduce the rotational 
and magnetic broadening effects on the Stokes profiles in wavelength space,
much as they are for the MZSs in Doppler space. We have tested our approach for 
different moderate values of $v\,\sin i$, 
obtaining good results in all cases. Gratifyingly enough it has turned out 
that the results do not depend on whether one uses fixed grids or adaptive grids.

{\sc Cossam}, the polarised spectral line synthesis code employed in this work, 
is considered a reference for the calculation of stellar Stokes parameters 
(\citealt{Wadeetal2001}, \citealt{CarrollKopStr2008}). For the inversion -- 
in the general case -- of the stellar MZSs, we adopted fixed stellar atmospheric 
model parameters ($T_{\rm eff}$, $log\,g$, $[M/H]$, $v_{\rm turb}$, $\xi$), but we 
freely varied the parameters of the magnetic model incorporated in {\sc Cossam}, 
viz. the magnetic moment $m$, the Eulerian angles
$\alpha$, $\beta$, $\gamma$, the vector of the dipole offset [0, $x_2$, $x_3$] and
the inclination $i$. We were able to demonstrate that ``solar'' broadened MZSs are
able to reproduce the shape of stellar MZSs, providing us with the possibility to
determine $H_{\rm eff}$. Currently, {\sc Cossam} assumes a tilted eccentric dipole but
our technique should also work for quadrupolar or higher order configurations, even
for magnetic fields concentrated in starspots.

 Computing times for the (point-source) ``solar'' MZSs are of course largely 
inferior to those for (integrated) stellar MZSs which increase proportionally to 
the number of spatial quadrature points. Still, for the approach presented here 
extensive calculations are required to arrive at a significant number of stellar 
MZSs needed to establish the BFs that serve to broaden the ``solar'' MZSs. About 3 
weeks had to be spent on a 56 core workstation to obtain 7500 stellar MZSs covering 
the interval from 350 to 1000\,nm in steps of 1\,km\,s$^{-1}$, and adopting an 80 
point spatial grid. It might at this point legitimately be asked why one should 
spend such an effort on the determination of $H_{\rm eff}$ instead of trying to 
directly model the stellar MZSs. Given the huge number of combinations in the
parameter space of tilted eccentric dipole model, an obvious 
reason for wanting to know the effective longitudinal field at various phases is 
to reduce this number, possibly by a very large amount. The availability of
(even modestly sized) supercomputers can greatly increase the attractiveness of 
our method. In a forthcoming paper we shall present the application of this 
technique to observational data.

\section*{Acknowledgements}
JRV, SGN and LS acknowledge support from the {\sc CONACyT} grants 240441, 168078
and 180817 respectively. Part of the results presented here have been obtained using 
the {\it ``Supercomputo - DGTIC''} facilities, grant SC16-1-IR-40,  of the UNAM, the 
computers from the {\sc CONACyT} project 153985 and the UNAM-PAPIIT project 107215,  
and the computers ``Tychos'' (Posgrado en Astrofisica-UNAM, 
Instituto de Astronomia-UNAM and {\sc PNPC-CONACyT}). 
LS also acknowledges support from the grant UNAM-DGAPA-PAPIIT  IA101316.
Thanks go to AdaCore for providing the GNAT GPL Edition of its Ada compiler.

\bibliographystyle{aa}
\bibliography{biblio}

\end{document}